\documentclass[a4paper,11pt]{article}
\usepackage[utf8]{inputenc}

\usepackage[colorlinks=true,linkcolor=blue, citecolor=red, urlcolor=blue, bookmarks]{hyperref} 
\usepackage[text={17.1cm,24.6cm},centering]{geometry} 

\usepackage[normalem]{ulem}

\usepackage{physics}  
\usepackage{amsfonts}  
\usepackage{amsmath}

\usepackage{tikz} 
\usepackage{graphicx}
\usepackage{subcaption}
\usepackage{caption}
\usepackage{xcolor}
\usepackage{ifthen}

\usepackage{xfp}
\usepackage[space]{cite}
\usepackage{dutchcal}

\newcommand{\id}{\mathbb{I}}
\newcommand{\Nat}{\mathbb{N}}
\newcommand{\non}{\nonumber}
\newcommand{\setF}{\mathcal{F}}
\newcommand{\setA}{\mathcal{A}}
\newcommand{\rrangle}{\rangle\!\rangle}
\newcommand{\llangle}{\langle\!\langle}

\numberwithin{equation}{section}

\usepackage{authblk}

\begin{document}
\title{\bf Entanglement of inhomogeneous free fermions on hyperplane lattices}
\author[1]{Pierre-Antoine Bernard}
\author[2]{Nicolas Cramp\'e}
\author[3]{Rafael I. Nepomechie}
\author[1]{Gilles Parez \footnote{ {\normalsize 
\texttt{gilles.parez@umontreal.ca}}}}
\author[4,5]{Lo\"ic~Poulain~d'Andecy}
\author[1,6]{Luc Vinet}
\affil[1]{\it Centre de Recherches Math\'ematiques, Universit\'e de Montr\'eal, P.O. Box 6128, Centre-ville
Station, Montr\'eal (Qu\'ebec), H3C 3J7, Canada}
\affil[2]{\it Institut Denis-Poisson CNRS/UMR 7013, Universit\'e de Tours - Université d’Orl\'eans, Parc de Grandmont, 37200 Tours, France}
\affil[3]{\it Physics Department, P.O. Box 248046, University of Miami, Coral Gables, FL 33124, USA}
\affil[4]{\it Laboratoire de Math\'ematiques de Reims UMR 9008, Universit\'e de Reims Champagne-Ardenne, Moulin de la Housse BP 1039, 51100 Reims, France}
\affil[5]{\it IRL-CRM, CNRS UMI 3457, Universit\'e de Montr\'eal, Canada}
\affil[6]{\it IVADO, 6666 Rue Saint-Urbain, Montr\'eal (Qu\'ebec), H2S 3H1, Canada}
\date{June 13, 2022}
\maketitle

\begin{abstract}

    We introduce an inhomogeneous model of free fermions on a $(D-1)$-dimensional lattice with $D(D-1)/2$ continuous parameters that control the hopping strength between adjacent sites. We solve this model exactly, and find that the eigenfunctions are given by multidimensional generalizations of Krawtchouk polynomials. We construct a Heun operator that commutes with the chopped correlation matrix, and compute the entanglement entropy numerically for $D=2,3,4$, for a wide range of parameters. For $D=2$, we observe oscillations in the sub-leading contribution to the entanglement entropy, for which we conjecture an exact expression. For $D>2$, we find logarithmic violations of the area law for the entanglement entropy with nontrivial dependence on the parameters.
\end{abstract}

\newpage

\tableofcontents

\section{Introduction}

This paper studies the entanglement entropy of free fermions on 
inhomogeneous lattices in various dimensions.

In quantum many-body physics, bipartite entanglement relates to how much a
part denoted $\setA$ of a system is correlated with its complement $\bar{\setA}$.
Quantifying this property is naturally of interest and warrants systematic investigation
\cite{AFOV08,ECP08, Latorre:2009zz} as it plays in particular a key role in
the characterization of critical points \cite{OAFF02,ON02,VLRK03,CC04} and
is of relevance in quantum information \cite{nielsen2002quantum}. One such way of quantifying
entanglement is through the entanglement entropy, provided by the von Neumann
entropy $S_{vN}$ of the reduced density matrix $\rho_\setA$ of $\setA$. For a
system in the pure state $| \Psi \rrangle$, the entanglement entropy is
defined by
\begin{equation}
S_{vN} = - \Tr(\rho_\setA \log \rho_\setA), \quad  \quad \rho_\setA
= \Tr_{\bar \setA}| \Psi \rrangle \llangle \Psi |.
\label{EE}
\end{equation}

The entanglement entropy computation for an $N$-body system in the
large-$N$ limit is generally a challenging problem.  Considerable attention
has been focused on free-fermion models in one dimension, where this
computation becomes tractable. Indeed, owing to the 
quadratic nature of such models, the spectrum of the reduced density
matrix can be obtained solely from the truncated two-point correlation
matrix $C$ \cite{peschel2003calculation}, 
which readily yields the entanglement entropy. For one-dimensional homogeneous chains, analytic results are available, see e.g. \cite{jin2004,its2005entanglement, calabrese2010universal, fagotti2011universal,eisler2013free, eisler2017analytical, eisler2018properties}. In particular, for the spin-1/2 XX spin chain with open boundary conditions, exact calculations based on the generalized Fisher-Hartwig conjecture \cite{fagotti2011universal} show that the entanglement entropy of a block of $\ell$ contiguous sites adjacent to a boundary scales as  
\begin{equation}
\label{eq:SvnScaling}
    S_{vN} = \frac 16 \log \ell + \dots \, 
\end{equation}
in the limit of a large system embedded in an infinite chain. This is in agreement with results of conformal field theory (CFT) \cite{holzhey1994geometric,CC04,CC05} with central charge $c=1$.

Over recent years, a strong theoretical effort has also been aimed at better understanding entanglement properties of spatially-inhomogeneous systems \cite{ramirez2015entanglement,DSVC17,rodriguez2017more,Crampe:2019upj,Crampe:2021,FA20,BSR21,FA21}. While analytical methods used for homogeneous chains do not generalize to the inhomogeneous case, results have been obtained using CFT in curved background \cite{DSVC17,rodriguez2017more,FA21}. In Ref. \cite{FA21}, the authors use this formalism to argue that the entanglement entropy of a large class of inhomogeneous chains scales as in Eq. \eqref{eq:SvnScaling}, where the inhomogeneous nature of the chains affects the sub-leading terms.

These one-dimensional models, both homogeneous and inhomogeneous, have a
deep connection to orthogonal polynomials (more precisely, families of
uni-variate orthogonal polynomials that belong to the Askey scheme \cite{koekoek2010hypergeometric}).
An additional interesting feature of these models, made possible by the
underlying bispectral context, is the existence of a tridiagonal Heun operator that commutes with $C$.
The identification of this commuting operator exploits the parallel 
between the diagonalization of $C$ (which is constructed out of 
operators projecting onto (i) the set of sites forming part $\setA$ and (ii) 
states corresponding to energies that are present in the Fermi sea) with discrete-discrete
problems in signal processing (that look at optimizing the concentration in
time of a band-limited function), see \cite{grunbaum2018algebraic}. 

Motivated by these results, we introduce here an inhomogeneous model of
free fermions that hop on a $(D-1)$-dimensional hyperplane sublattice of a
$D$-dimensional hypercubic lattice. This model has $D(D-1)/2$ continuous parameters
that control the strength of hopping between nearest-neighbor sites, as 
well as $D$ discrete parameters. For
$D=2$, this model reduces to the Krawtchouk chain \cite{Crampe:2019upj}.  We solve the general
model exactly, and find that the eigenfunctions are given by
multidimensional generalizations of Krawtchouk polynomials. These polynomials were introduced by Griffiths more than 50 years ago \cite{griffiths1972orthogonal}, rediscovered at the turn of the millennium
and much studied thereafter \cite{ mizukawa2004n+, hoare2008probablistic, iliev2012rahman, iliev2012lie, diaconis2014introduction,Iliev2007,grunbaum2011system,geronimo2010bispectrality,GVZ2103}. For simplicity, we 
focus here on special cases (with one or more parameters set to zero) that were 
studied by Tratnik \cite{tratnik1991some}\footnote{These
multivariate polynomials of Griffiths and Tratnik types have also been used
recently to design spin lattices with useful properties such as perfect
state transfer \cite{miki2012quantum, miki2019quantum}.}. Moreover, we construct a multidimensional generalization of the $D=2$ Heun operator 
found in \cite{Crampe:2019upj} which commutes with the chopped correlation matrix $C$.

We compute the entanglement entropy numerically for $D=2,3,4$, for a wide range of different parameters. For $D=2$, where the model has just one continuous parameter $p \in (0,1)$, we confirm recent results for $p=1/2$ \cite{FA21}, and obtain new results for $p\ne 1/2$. In particular, we observe oscillations in the sub-leading contribution to the entanglement entropy, for which we conjecture an exact expression.
For $D>2$, we find logarithmic violations of the area law for the entanglement entropy with nontrivial dependence on the parameters.  

The paper will unfold as follows. The multidimensional inhomogeneous free-fermion model is introduced in Sec. \ref{sec:model}, and various special cases are spelled out.  Their analytic solutions are provided. A quick review of the characterization of the Griffiths polynomials is also given.  The framework for the entanglement study is established in Sec. \ref{sec:cormat} by specifying the eigenstate in which the system will be taken, and by formulating the correlation matrix in that state.  Upon fixing what has been called the part $\setA$ of the system (the part whose entanglement we wish to concentrate on), the appropriate multidimensional generalization of the Heun operator is constructed and shown to commute with the chopped correlation matrix. This sets the stage for the determination of the entanglement entropy of various cases which is carried out in Sec. \ref{sec:entanglement}. Conclusions and perspectives are offered in Sec. \ref{sec:end}.

\section{The inhomogeneous free-fermion model}\label{sec:model}

In this section we define an inhomogeneous free-fermion model on a hyperplane lattice, and derive its exact solution. We introduce the model in Sec. \ref{sec:defH}, and point out special cases for $D>2$, namely the Tratnik and the one-parameter cases. We diagonalize the general Hamiltonian in Sec. \ref{sec:Solution} with the help of eigenfunctions that are investigated in more detail in Sec. \ref{sec:eigenfunctions}. In particular, we obtain closed-form formulas for the various special cases under consideration in terms of Krawtchouk polynomials, which will be used for the computations in Sec. \ref{sec:entanglement}.

\subsection{Definition of the model}\label{sec:defH}

We consider free fermions defined on the set of points $V =\{ \vec x \in 
\Nat^{D} | x_{1} + \ldots + x_{D} = N\,, x_{i} \ge 0 \}$, that is, 
vertices of a hypercubic lattice in $D$ dimensions satisfying the constraint
\begin{equation}
	\sum_{i=1}^{D} x_{i} = N\,,
\label{xconstraint}	
\end{equation}
which defines a $(D-1)$-dimensional hyperplane. The number of such vertices is
\begin{equation}
	n(N,D) =  {N+D-1\choose D-1} \,.
\label{dim}
\end{equation}
 The parameter $N$ gives the diameter of the hyperplane, and two sites $\Vec{x}$ and $\Vec{y}$ are nearest neighbors if there exist distinct indices $i, j \in \{1, \dots, D\}$ such that 
\begin{equation}
    \Vec{x} - \Vec{y} =  \Vec{\epsilon}_i - \Vec{\epsilon}_j,
\end{equation}
where 
\begin{align}
     \Vec{\epsilon}_i = (\underbrace{0, \dots, 0 }_{i-1 \text{ times}}, 1,\underbrace{0, \dots, 0 }_{D- i \text{ times}}). 
     \label{epsi}
\end{align}

\begin{figure}
    \centering
     \includegraphics[width=.4\linewidth]{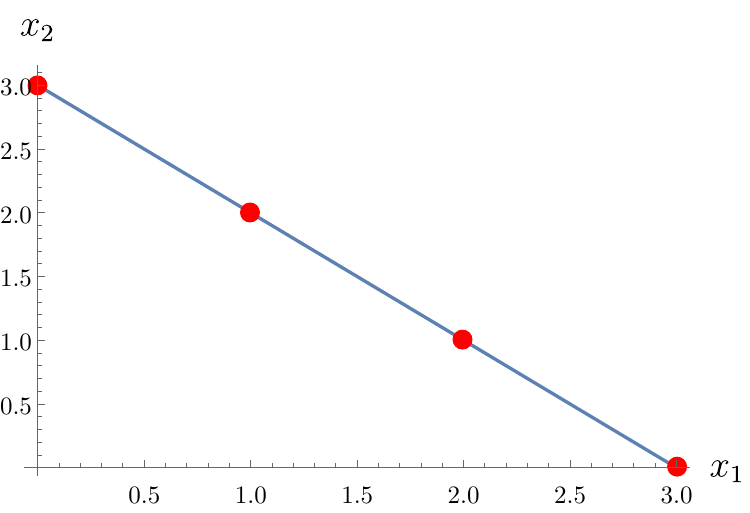}
      \includegraphics[width=.4\linewidth]{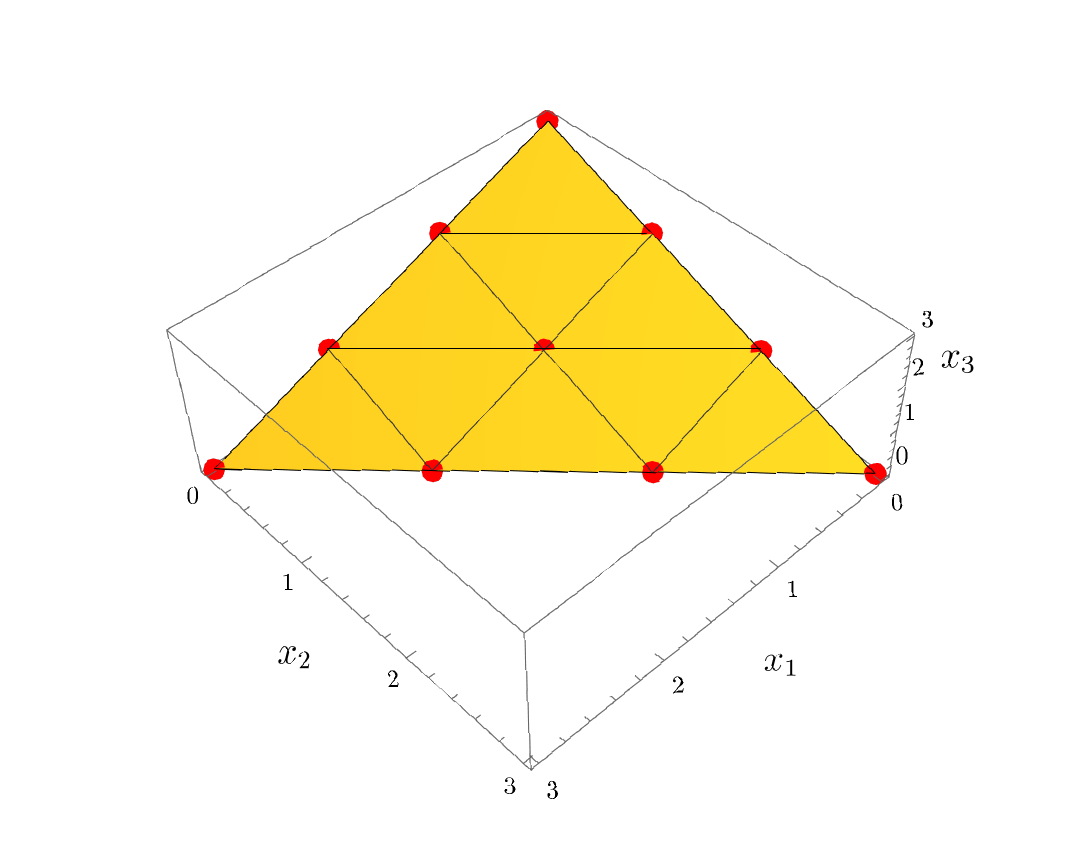}
    \caption{{\it Left}: Lattice for $D=2, \ N=3$. {\it Right:} Lattice for $D=3, \ N=3$.}
    \label{fig:D23}
\end{figure}

As examples of the hyperplane lattices we consider, the case $D=2$ corresponds to a linear lattice, and 
the case $D=3$ corresponds to a planar triangular lattice, as illustrated in Fig. \ref{fig:D23}. In these figures, the vertices are denoted by dots, and nearest neighbors are connected by lines. The maximum possible number of nearest neighbors that a lattice site can have in these examples is 2 and 6, respectively, and in general is given by $D(D-1)$. In contrast, on a $D$-dimensional hypercubic lattice without the constraint \eqref{xconstraint}, the maximum number of nearest neighbors is only $2D$. For $D=4$, the corresponding lattice is a tetrahedron obtained as the superposition of planar triangular lattices, but we do not represent it here.

The Hamiltonian is given by
\begin{equation}
	{\cal H} = \sum_{\vec x \in V} \sum_{i=1}^{D} \alpha_{i} \left\{
\sum_{j,k=1; j\ne k}^{D} R_{ij}\, R_{ik} \sqrt{x_{j} (x_{k}+1)} 
c^{\dagger}_{\vec x - \vec\epsilon_{j}+ \vec\epsilon_{k}}\, c_{\vec x} 
+ \sum_{j=1}^{D} R_{ij}^{2}\, x_{j}\, c^{\dagger}_{\vec x}\, c_{\vec x} \right\} \,,
\label{Hamiltonian}
\end{equation}
where the fermion annihilation and creation operators obey the usual 
anticommutation relations
\begin{equation}
\left\{ c_{\vec x'} \,,  c_{\vec x} \right\} = 0 = \left\{ 
c^{\dagger}_{\vec x'} \,, c^{\dagger}_{\vec x} \right\}\,,  \qquad
\left\{ c_{\vec x'} \,, c^{\dagger}_{\vec x} \right\} = \delta_{\vec 
x',\vec x} \,,
\end{equation}
and $R$ is an $SO(D)$ matrix, namely
\begin{equation}
R^{t}\, R = R\, R^{t} =\id\,, \qquad \det(R) = 1\ .
\end{equation}

In the Hamiltonian \eqref{Hamiltonian}, for each value of $i$, one has the same types of hopping between nearest neighbors but with different coupling parameters. However, these parameters are related by virtue of being elements of a rotation matrix $R$. This matrix can be expressed as $R=e^{B}$, where $B$ is a real antisymmetric $D \times D$ matrix which has $D(D-1)/2$ independent parameters. Finally, we restrict the real parameters $\vec \alpha = (\alpha_{1}, \ldots, \alpha_{D})$ to discrete values $\alpha_{i} \in \{0, 1 \}$, which is required for the construction of the Heun operator in Sec. \ref{sec:Heun}. The first term in the Hamiltonian \eqref{Hamiltonian} corresponds to nonuniform nearest-neighbor hopping, while the  second term corresponds to a nonuniform chemical potential. In the following, we choose $R$ such that \eqref{Hamiltonian} reduces to known inhomogeneous models in one dimension (for $D=2$) and gives solvable generalizations thereof in higher dimensions.

\subsubsection[The case $D=2$]{The case $\boldsymbol{D=2}$}

For $D=2$, with $\vec x = (x, N-x)$, $\vec \alpha 
=(1,0)$, $p\in \mathbb{R}$ such that $0 < p < 1$ and
\begin{equation}
R = 
\begin{pmatrix}
	\sqrt{1-p} & \sqrt{p} \\
	-\sqrt{p} & \sqrt{1-p} 
	\end{pmatrix}\,,
\label{RKraw}
\end{equation}	 
the Hamiltonian \eqref{Hamiltonian} becomes
\begin{align}
	{\cal H} = \sum_{x=0}^{N-1} 
\sqrt{p(1-p)(x+1)(N-x)}\left( c^{\dagger}_{x}\, c_{x+1} 
+c^{\dagger}_{x + 1}\, c_{x} \right)  + \sum_{x=0}^{N} \left(p N + x(1-2p)\right)\, c^{\dagger}_{x}\, c_{x} \,.
\label{HD2}
\end{align}
We recognize this model as one associated
with the Krawtchouk polynomials, see e.g. \cite{Crampe:2019upj}.  For general values of $D$, 
we shall see that the Hamiltonian \eqref{Hamiltonian} is similarly
associated with multivariate Krawtchouk polynomials.

\subsubsection{Tratnik cases}

The case $D=3$ with $R_{12}=0$ is associated with the so-called
bivariate Krawtchouk polynomials of Tratnik type \cite{tratnik1991some, geronimo2010bispectrality, GVZ2103}. For this case, the matrix  
$R$ depends on 2 (rather than 3) independent parameters, which we denote by 
$p_{1}=R_{13}^2$ and $p_{2}=R_{23}^2$:
\begin{equation}
R= \begin{pmatrix}
\sqrt{1-p_{1}} 	&0  &\sqrt{p_{1}} \\
 -\sqrt{\frac{p_{1} p_{2}}{1-p_{1}}} 
& \sqrt{\frac{1-p_{1} -p_{2}}{1-p_{1}}}
& \sqrt{p_{2}}\\
-\sqrt{\frac{p_{1}(1-p_{1}- p_{2})}{1-p_{1}}}
& -\sqrt{\frac{p_{2}}{1-p_{1}}}
&\sqrt{1-p_{1}-p_{2}}
\end{pmatrix}\,.
\label{Rtratnik}
\end{equation}

Similarly, for  $D=4$, the Tratnik-type rotation matrix has 3 zero matrix elements, and 3 independent parameters that we denote by $p_{1}=R_{14}^2$, $p_{2}=R_{24}^2$ and $p_{3}=R_{34}^2$:
\begin{equation}
R= \begin{pmatrix}
\sqrt{1-p_1} & 0 & 0 & \sqrt{p_1} \\
-\sqrt{\frac{p_1 p_2}{1-p_1}} & \sqrt{\frac{1-p_1-p_2}{1-p_1}} &0 & \sqrt{p_2}\\
\sqrt{\frac{p_1 p_3}{1-p_1}} & \sqrt{\frac{p_2 p_3}{(1-p_1)(1-p_1-p_2)}} & \sqrt{\frac{1-p_1-p_2-p_3}{1-p_1-p_2}} & -\sqrt{p_3} \\
-\sqrt{\frac{p_1(1-p_1-p_2-p_3)}{1-p_1}} & -\sqrt{\frac{p_2(1-p_1-p_2-p_3)}{(1-p_1)(1-p_1-p_2)}} & \sqrt{\frac{p_3}{1-p_1-p_2}} & \sqrt{1-p_1-p_2-p_3} 
\end{pmatrix}\,.
\label{Rtratnik4}
\end{equation}

\begin{figure}[h]
\begin{subfigure}{.5\textwidth}
  \centering
  \includegraphics[width=.8\linewidth]{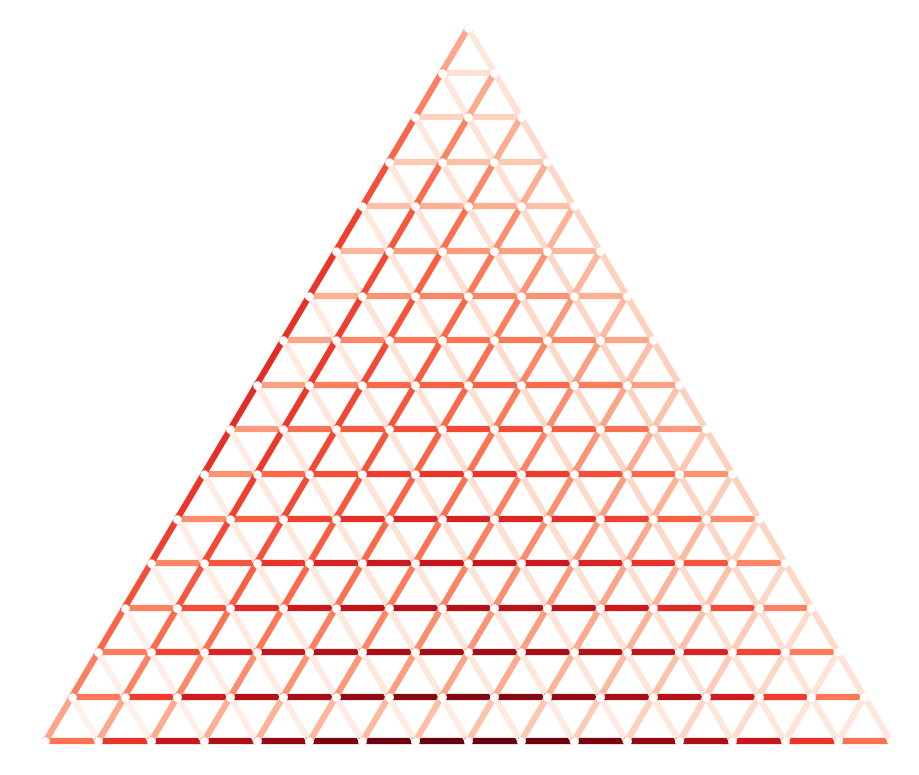}
  \caption{$p_1 = 1/32$, $p_2 = 1/16$, $\vec{\alpha} = (1,1,0)$}
  \label{fig:sfig1}
\end{subfigure}%
\begin{subfigure}{.5\textwidth}
  \centering
  \includegraphics[width=.8\linewidth]{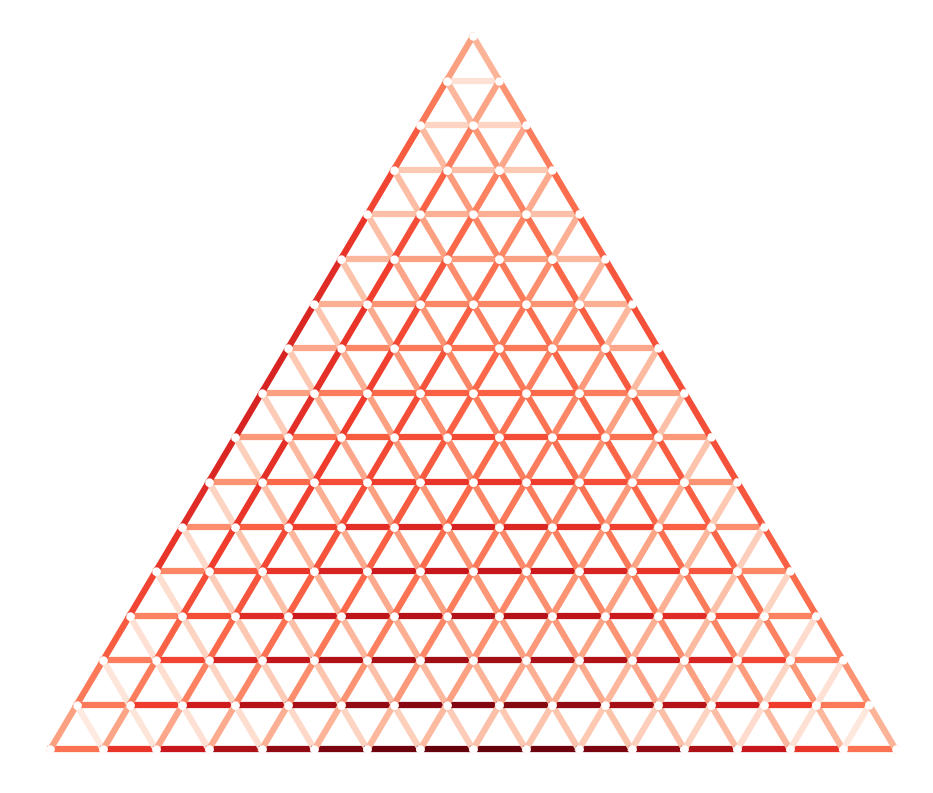}
  \caption{$p_1 = 1/4$,  $p_2 = 1/2$, $ \vec{\alpha} = (0,1,0)$}
  \label{fig:sfig2}
\end{subfigure}
\caption{Examples of inhomogeneous lattices corresponding to the Tratnik case for $D = 3$ and $N = 16$. The color magnitude represents the strength of the couplings between neighboring sites, and darker colors correspond to stronger interactions.}
\label{fig:trat}
\end{figure}

In Fig. \ref{fig:trat} we illustrate the lattice corresponding to the Tratnik case for $D=3$ for different values of $p_1,p_2$ and $\vec \alpha$. The color on each edge represents the magnitude of the hopping interaction between the two respective adjacent sites in the Hamiltonian \eqref{Hamiltonian}. The actual values of the couplings are irrelevant, but darker colors represent stronger hopping interactions. The inhomogeneous nature of the model is clear.

\subsubsection{One-parameter cases}

The $D=3$ rotation matrix with a single parameter $p$,
\begin{equation}
R= \begin{pmatrix}
1 	&0  &0 \\
0 & \sqrt{1-p} & \sqrt{p}\\
0 & -\sqrt{p} & \sqrt{1-p}
\end{pmatrix},
\label{Roneparam}
\end{equation}
is associated with a single uni-variate Krawtchouk polynomial. This matrix can be obtained from \eqref{Rtratnik} by setting $p_1=0$ and  $p_2 = p$.

Similarly, the $D=4$ rotation matrix \eqref{Rtratnik4} reduces for $p_1=p_2=0$ and $p_3 = p$ to 
\begin{equation}
R= \begin{pmatrix}
1 & 0 & 0 & 0 \\
0 & 1 & 0 & 0 \\
0 & 0 & \sqrt{1-p} & -\sqrt{p} \\
0 & 0 & \sqrt{p} & \sqrt{1-p} 
\end{pmatrix}\,,
\label{RoneparamD4}
\end{equation}
and is also associated with a single uni-variate Krawtchouk polynomial.
	
\subsection{Solution}\label{sec:Solution}

In order to solve this model, we begin by rewriting the Hamiltonian \eqref{Hamiltonian} 
in terms of a set of $n(N,D) \times n(N,D)$ matrices $\left( H_{i} \right)_{\vec x',\vec x}$,
\begin{equation}
{\cal H} = \sum_{\vec x, \vec x' \in V} \sum_{i=1}^{D} \alpha_{i}\, 
 c^{\dagger}_{\vec x'}\, \left( H_{i} \right)_{\vec x',\vec x}\, c_{\vec x}  \,,
\label{Hamiltonian2}
\end{equation}
where
\begin{equation}
\left( H_{i} \right)_{\vec x',\vec x} = \sum_{j,k =1; j \ne k}^{D} 
R_{ij} R_{ik}\,\sqrt{x_{j} (x_{k}+1)}
\delta_{\vec x', \vec x - \vec\epsilon_{j} + \vec\epsilon_{k}}
+ \sum_{j=1}^{D} R_{ij}^{2} x_{j} \delta_{\vec x', \vec x} \,.
\label{Hxx}
\end{equation}

We now proceed to diagonalize the matrices $\left( H_{i} \right)_{\vec x',\vec x}$. To this end, let 
us introduce, following \cite{GVZ2103}, a set of $D$ harmonic oscillator operators with commutation 
relations
\begin{equation}
\left[a_{i} \,, a_{j} \right] = 0 \,, \qquad 
\left[a^{\dagger}_{i} \,, a^{\dagger}_{j} \right] = 0 \,,  \qquad 
\left[a_{i} \,, a^{\dagger}_{j} \right] = \delta_{i,j} \,,  \qquad
i, j = 1, \ldots, D \,.
\label{hoalgebra}
\end{equation}
The algebra \eqref{hoalgebra} has a representation on the orthonormal states 
\begin{equation}
|\vec x \rangle = |x_{1} \rangle \otimes \ldots \otimes |x_{D} 
\rangle \,, \qquad 
\vec x \in \Nat^{D} \,, \qquad x_{i} \ge 0 \,,
\end{equation}
defined by
\begin{align}
a_{i} | \vec x \rangle &= \sqrt{x_{i}} | \cdots, x_{i}-1, \cdots 
\rangle \,, \non \\
a^{\dagger}_{i} | \vec x \rangle &= \sqrt{x_{i}+1} | \cdots, x_{i}+1, \cdots 
\rangle \,.
\label{arep}
\end{align}

We define the operators $X_{i}$ and $H_{i}$ in terms of the harmonic 
oscillator operators by 
\begin{equation}
X_{i} = a^{\dagger}_{i}\, a_{i} \,, \qquad 
H_{i} = \sum_{j,k=1}^{D}R_{ij}\, R_{ik}\, a^{\dagger}_{j}\, a_{k}\,, 
\qquad i = 	1, \ldots, D \,,
\label{XH}
\end{equation}
which satisfy\footnote{We also remark that
\begin{align}
	\left[ X_{i} \,, \left[ X_{i} \,, \left[ X_{i} \,, H_{j} \right]  
	\right]  \right]  &= \left[ X_{i} \,, H_{j} \right] \,, \non \\
	\left[ H_{i} \,, \left[ H_{i} \,, \left[ H_{i} \,, X_{j}\right]  
	\right]  \right]  &= \left[  H_{i} \,, X_{j} \right] \,. \non
\end{align}
Moreover, for generic parameters $R$, the two sets of operators $X_i$ and $H_i$ generate together the whole $su(D)$-algebra realized by the operators $a^{\dagger}_{j}\, a_{k}$. Indeed, we have
\[\frac{1}{2R_{ij}R_{ik}}\bigl([X_k,[X_j,H_i]]-[X_j,[X_k,[X_j,H_i]]]\bigr)=a^{\dagger}_{j}\, a_{k}\,,\]
see also Lemma 2.2 in \cite{iliev2012lie}.}
\begin{equation}
\left[X_{i}\,, X_{j} \right] = 0 \,, \qquad  \left[H_{i}\,, H_{j} \right]  = 0 \,.
\end{equation}	

It is important to note that these operators are related by a unitary transformation,
\begin{equation}
U(R)^{\dagger}\, X_{i}\, U(R)  = H_{i}\,, \qquad i = 	1, \ldots, D\,,
\label{XHreltn}
\end{equation}	
where $U(R)$ is defined by \cite{GVZ2103}
\begin{equation}
	U(R) = \exp\left(\sum_{k,l=1}^{D} B_{kl}\, a^{\dagger}_{k} a_{l} \right) \,,
\label{UR}
\end{equation}	
and whose action on the harmonic oscillator operators is
\begin{equation}
U(R)\, a^{\dagger}_{i}\, U(R)^{\dagger} = \sum_{j=1}^{D} R_{ji}\, 
a^{\dagger}_{j} \,, \qquad 
U(R)\, a_{i}\, U(R)^{\dagger} = \sum_{j=1}^{D} R_{ji}\, 
a_{j} \,.
\end{equation}	

In view of \eqref{arep}, we see that
\begin{equation}
X_{i} | \vec x \rangle = x_{i}  | \vec x \rangle \,, \qquad i = 1, \ldots, D\,,
\label{Xxbasis}
\end{equation}
and 
\begin{equation}
H_{i} | \vec x \rangle = \sum_{j,k =1; j \ne k}^{D} 
R_{ij} R_{ik}\,\sqrt{x_{j} (x_{k}+1)}| \vec x - \vec\epsilon_{j} + \vec\epsilon_{k}\rangle 
+ \sum_{j=1}^{D} R_{ij}^{2} x_{j} | \vec x \rangle \,.
\label{Hax}
\end{equation}
For a given $N$, we restrict ourselves to values of $\vec x$ that satisfy \eqref{xconstraint}.
The result \eqref{Hax} implies that the matrices \eqref{Hxx} can be expressed as matrix elements of $H_{i}$,
\begin{equation}
\left( H_{i} \right)_{\vec x',\vec x} = \langle \vec x' | H_{i} | \vec x \rangle \,.
\label{Hxbasis}
\end{equation}

From \eqref{XHreltn} and \eqref{Xxbasis}, we see that
\begin{equation}
H_{i} | \vec k \rangle = k_{i}  | \vec k \rangle \,, \qquad i = 1, \ldots, D\,,
\label{Hkbasis}
\end{equation}
where the eigenvectors $| \vec k \rangle$ are given by
\begin{equation}
| \vec k \rangle =	U(R)^{\dagger}\, | \vec x \rangle \Big\vert_{\vec x=\vec k}\,,
\label{kbasis}
\end{equation}
and the corresponding eigenvalues are given by $\vec k = \vec x \in 
V$. In particular, $\sum_{i=1}^{D} k_{i} = N$.
We note that $\langle \vec x |  \vec k \rangle = \langle \vec k 
|  \vec x \rangle$, and
\begin{equation}
\sum_{\vec k \in V} \langle \vec x' |  \vec k \rangle 
\langle \vec k |\vec x \rangle = \delta_{\vec x', \vec x} \,, \qquad
\sum_{\vec x \in V} \langle \vec k' |  \vec x \rangle 
\langle \vec x |\vec k \rangle = \delta_{\vec k', \vec k} \,.
\end{equation}

We can also obtain the result
\begin{equation}
X_{i}\, | \vec k \rangle = \sum_{j,l =1; j \ne l}^{D} 
R_{ji} R_{li}\, \sqrt{k_{j}(k_{l}+1)}\, 
| \vec k  - \vec\epsilon_{j} + \vec\epsilon_{l} \rangle  
+ \sum_{j=1}^{D} R_{ji}^{2} k_{j}\, | \vec k \rangle \,
\label{Xak}
\end{equation}
by acting with  $H'_{i}  \equiv \sum_{j,k=1}^{D}R_{ji}\, R_{ki}\, a^{\dagger}_{j}\, a_{k}$ on $| \vec x \rangle$, obtaining a result similar to \eqref{Hax}, and then using $H'_{i} = U(R)\, X_{i}\, U(R)^{\dagger}$
and \eqref{kbasis}. 
We see from Eqs. \eqref{Xxbasis}, \eqref{Hax}, \eqref{Hkbasis} and
\eqref{Xak} that the overlaps~$\langle \vec x |  \vec k \rangle$ are solutions
to a bispectral problem defined from the set of equations obtained by equating in
$\langle \vec x | X_i |\vec k \rangle$ and $\langle \vec x | H_i | \vec k \rangle$
the actions of $X_i$ and $H_i$ respectively on the bra and on the ket
\footnote{Recall that a function of two sets of variables is said to be solution of a bispectral problem \cite{duistermaat1986differential}
if it is an eigenfunction of operators acting on the variables of the first set with eigenvalues
depending on the variables of the second set and vice versa. With their recurrence relations and their differential or difference equation,
classical orthogonal polynomials are standard examples of such solutions.}.

In terms of the overlaps, the Hamiltonian \eqref{Hamiltonian} reads
\begin{align}
{\cal H} &= \sum_{\vec x, \vec x' \in V} 
 c^{\dagger}_{\vec x'}\, \langle \vec x'| \vec\alpha \cdot \vec H | 
 \vec x \rangle \, c_{\vec x}  \non\\
 &= \sum_{\vec x, \vec x', \vec k \in V} 
 c^{\dagger}_{\vec x'}\, \langle \vec x'| \vec\alpha \cdot \vec H | 
 \vec k \rangle \langle \vec k | \vec x \rangle \, c_{\vec x} \non\\
 &= \sum_{\vec x, \vec x', \vec k \in V} 
 \vec\alpha \cdot \vec k \,
 \langle \vec x' |  \vec k \rangle 
 \langle \vec k | \vec x \rangle \, 
 c^{\dagger}_{\vec x'}\,  c_{\vec x}
 \,,
\label{Hamiltonian3}
\end{align}
where we used \eqref{Hamiltonian2}, \eqref{Hxbasis} and 
\eqref{Hkbasis}. We introduce Fourier-transformed fermionic operators~$\tilde{c}_{\vec k}$,
\begin{equation}
\tilde{c}_{\vec k} = \sum_{\vec x \in V} \langle \vec x |  \vec k \rangle 
c_{\vec x} \,, \qquad
c_{\vec x} = \sum_{\vec k \in V} \langle \vec x |  \vec k \rangle 
\tilde{c}_{\vec k} \,,
\end{equation}
which satisfy the anticommutation relations
\begin{equation}
\left\{ \tilde{c}_{\vec k'} \,,  \tilde{c}_{\vec k} \right\} = 0 = 
\left\{ \tilde{c}^{\dagger}_{\vec k'} \,, \tilde{c}^{\dagger}_{\vec k} \right\}\,,  \qquad
\left\{ \tilde{c}_{\vec k'} \,, \tilde{c}^{\dagger}_{\vec k} \right\} = 
\delta_{\vec k',\vec k} \,.
\label{ctildeccr}
\end{equation}
Performing the sums over $\vec x$ and $\vec x'$ in  
\eqref{Hamiltonian3}, we finally arrive at the diagonal form of 
the Hamiltonian,
\begin{equation}
{\cal H} = 	\sum_{\vec k \in V} \vec\alpha \cdot \vec k\,   
\tilde{c}^{\dagger}_{\vec k}\,  \tilde{c}_{\vec k} \,.
\label{Hamiltonian4}
\end{equation}

The eigenvectors of the Hamiltonian are given by
\begin{equation}
|\Psi\rangle\!\rangle = \prod_{\vec k \in \setF} 
\tilde{c}^{\dagger}_{\vec k} |0\rangle\!\rangle \,,
\label{Fermioneigenstates}
\end{equation}
where the vectors in $\setF \subseteq V$ are pairwise distinct, and 
$|0\rangle\!\rangle$ is the vacuum state, 
\begin{equation}
\tilde{c}_{\vec k} |0\rangle\!\rangle = 0 \qquad \forall   \vec k \in V \,.
\end{equation}
Indeed,
\begin{equation}
{\cal H}\, |\Psi\rangle\!\rangle = E |\Psi\rangle\!\rangle \,, \qquad
E = \sum_{\vec k \in \setF}  \vec\alpha \cdot \vec k \,.
\label{spectrum}
\end{equation}
The number of fermions in the state $|\Psi\rangle\!\rangle$ \eqref{Fermioneigenstates}  is given by $|\setF|$, the cardinality of the set $\setF$.

We observe that the model's spectrum is integer valued, and does not
depend on the matrix $R$. This can be understood from the fact that
the Hamiltonian is constructed from operators $H_{i}$ that are related
to $X_{i}$ by a unitary transformation \eqref{XHreltn}. For the 
trivial rotation $R=\id$ (i.e., $R_{ij} = \delta_{i,j}$), 
we see that $H_{i}=X_{i}$, and the Hamiltonian \eqref{Hamiltonian} 
becomes diagonal in the $|\vec x\rangle$ basis,
\begin{equation}
{\cal H} = \sum_{\vec x \in V} \vec\alpha \cdot \vec x\, 
c^{\dagger}_{\vec x}\, c_{\vec x} \,, 
\end{equation}
which leads to the spectrum in \eqref{spectrum}. Although the 
spectrum does not depend on $R$, the eigenfunctions $\langle \vec x 
|\vec k \rangle$ do depend on $R$, and are quite nontrivial, as described below.

We conclude with two remarks regarding the spectrum. First, because the spectrum is integer valued, it appears to be gapped. However, because of the inhomogeneous nature of the model, one needs to investigate the gap in the scaling limit with respect to a rescaled system size, and for $D=2$ the system is in fact gapless \cite{FA21,BSR21}. Second, we note that the spectrum is highly degenerated. For a given $\vec \alpha$, the components of $\vec k$ that correspond to vanishing components of $\vec \alpha$ do not contribute to the energies and hence create the degeneracy. A refined understanding of the origin of the degeneracies and their physical implications is left for forthcoming investigations.

\subsection{Eigenfunctions}\label{sec:eigenfunctions}

We discuss here some further properties of the eigenfunctions $\langle \vec k 
|\vec x \rangle$, corresponding to the eigenvectors of the $H_{i}$ 
operators \eqref{Hkbasis}. We have 
\begin{equation}
\langle \vec k |\vec x \rangle = 
\langle \vec x'| U(R) |\vec x \rangle\Big\vert_{\vec x' = \vec k} 
= \sqrt{W(\vec x, \vec k)}\, Q_{\vec x}(\vec k)\,,
\label{overlap}
\end{equation}
where the first equality follows from \eqref{kbasis}, 
and the second equality follows from results in \cite{GVZ2103}.
The functions $Q_{\vec x}(\vec k)$ are multivariate Krawtchouk 
polynomials, which obey the recurrence relation \cite{GVZ2103}
\begin{equation}
k_{i} Q_{\vec x}(\vec k) = \sum_{j,l =1; j \ne k}^{D} 
\frac{R_{ij} R_{il} R_{D,k}}{R_{D,j}}\, x_{j}\,  
Q_{\vec x - \vec\epsilon_{j} + \vec\epsilon_{l}}(\vec k)
+ \sum_{j=1}^{D} R_{ij}^{2} x_{j} 
Q_{\vec x}(\vec k) \,, \qquad \vec x, \vec k \in V \,.
\label{recurrence}
\end{equation}
(While the notation of reference \cite{GVZ2103} will be adopted in the
following, the reader might wish to consult the following papers 
\cite{griffiths1972orthogonal, mizukawa2004n+, hoare2008probablistic, iliev2012rahman, iliev2012lie, diaconis2014introduction} for more on these polynomials.)
The polynomials  $Q_{\vec x}(\vec k)$ can be obtained from a generating function \cite{GVZ2103}
\begin{equation}
\prod_{i=1}^{D} \left(\sum_{j=1}^{D} \frac{R_{ij}R_{D,D}}{R_{i,D} R_{D,j}} 
z_{j}\right)^{k_{i}} = \sum_{\vec x \in V}\frac{N!}{x_{1}! \ldots x_{D}!}\, 
Q_{\vec x}(\vec k)\, z_{1}^{x_{1}} \ldots z_{D-1}^{x_{D-1}} \,, 
\label{genfunc}
\end{equation}
where $z_{D} \equiv 1$. Moreover, the function $W(\vec x, \vec k)$ in 
\eqref{overlap}, which is given by
\begin{equation}
W(\vec x, \vec k) =  \frac{N!}{x_{1}! \ldots x_{D}!} 
\frac{N!}{k_{1}! \ldots k_{D}!}  
\frac{\prod_{j=1}^{D} R_{D,j}^{2 x_{j}}}{R_{D,D}^{N}} 
\frac{\prod_{j=1}^{D} R_{j,D}^{2 k_{j}}}{R_{D,D}^{N}}\,,
\label{W}
\end{equation}
is the discrete measure appearing in the orthogonality relation \cite{GVZ2103},
\begin{equation}
\sum_{\vec k\in V}
W(\vec x, \vec k)\,
Q_{\vec x}(\vec k)\,
Q_{\vec x'}(\vec k) = \delta_{\vec x, \vec x'} \,.
\label{orthogonality}
\end{equation}

Let us perform a consistency check by using \eqref{overlap} to
rewrite the recurrence relation \eqref{recurrence} in terms of $\langle \vec k | \vec x \rangle$,
\begin{align}
k_{i} \langle \vec k | \vec x \rangle &= \sum_{j,l =1; j \ne l}^{D} 
\frac{R_{ij} R_{il} R_{D,l}}{R_{D,j}}\, x_{j} 
\sqrt{\frac{W(\vec x, \vec k)}{W(\vec x - \vec\epsilon_{j} + 
\vec\epsilon_{l}, \vec k)}}\,
\langle \vec k | \vec x - \vec\epsilon_{j} + \vec\epsilon_{l}\rangle  + \sum_{j=1}^{D} R_{ij}^{2} x_{j} 
\langle \vec k | \vec x \rangle \,.
\label{recurrence2}
\end{align}
Using \eqref{W}, we observe the following simplification,
\begin{equation}
\frac{W(\vec x, \vec k)}{W(\vec x - \vec\epsilon_{j} + \vec\epsilon_{l}, \vec k)}
= \frac{(x_{l}+1)}{x_{j}} \frac{R_{D,j}^{2}}{R_{D,l}^{2}} \,.
\end{equation}
Hence, the recurrence relation \eqref{recurrence2} takes the form
\begin{equation}
k_{i} \langle \vec k | \vec x \rangle = \sum_{j,l =1; j \ne l}^{D} 
R_{ij} R_{il}\,\sqrt{x_{j} (x_{l}+1)}
\langle \vec k | \vec x - \vec\epsilon_{j} + \vec\epsilon_{l}\rangle 
+ \sum_{j=1}^{D} R_{ij}^{2} x_{j} \langle \vec k | \vec x \rangle \,,
\label{recurrence3}
\end{equation}
which coincides with the result obtained by applying $\langle \vec k |$ 
to \eqref{Hax} and then using \eqref{Hkbasis}.

The functions $Q_{\vec x}(\vec k)$ also satisfy the difference equation \cite{GVZ2103}
\begin{equation}
	x_{i} Q_{\vec x}(\vec k) = \sum_{j,l =1; j \ne l}^{D} 
	\frac{R_{ji} R_{li} R_{l,D}}{R_{j,D}}\, k_{j}\, 
	Q_{\vec x}(\vec k - \vec\epsilon_{j} + \vec\epsilon_{l})
	+ \sum_{j=1}^{D} R_{ji}^{2} k_{j} Q_{\vec x}(\vec k) \,, \qquad \vec x, 
	\vec k \in V \,.
\label{diff}
\end{equation}
Similarly as before, we can use \eqref{overlap} to rewrite the difference equation in terms of 
$\langle \vec x | \vec k \rangle$, and obtain
\begin{equation}
x_{i} \langle \vec x | \vec k \rangle = \sum_{j,l =1; j \ne l}^{D} 
R_{ji} R_{li}\, \sqrt{k_{j}(k_{l}+1)}\, 
\langle \vec x | \vec k  - \vec\epsilon_{j} + \vec\epsilon_{l} \rangle  
+ \sum_{j=1}^{D} R_{ji}^{2} k_{j} \langle \vec x | \vec k \rangle \,,
\label{diff3}
\end{equation}
which coincides with the result obtained by applying $\langle \vec x |$ 
to \eqref{Xak} and then using \eqref{Xxbasis}.

\subsubsection[The case $D=2$]{The case $\boldsymbol{D=2}$}

For the case $D=2$ with the matrix $R$ given by \eqref{RKraw}, the 
functions $Q_{\vec x}(\vec k)$ can be expressed in terms of Krawtchouk 
polynomials,
\begin{equation}
Q_{x_{1}, x_{2}}(k_{1}, k_{2})=\frac{1}{(-N)_{x_{1}}} 
\mathcal{k}_{x_{1}}(k_{1};p;N) \,,
\label{QKrawt}
\end{equation}
where 
\begin{equation}
\mathcal{k}_{x}(k;p;N) = (-N)_{x}\ {}_{2}F_{1} \left(\begin{matrix} -k, & & -x
	\\&-N  &\end{matrix}\ 
	;\frac{1}{p}\right) \,,
\label{Krawt}
\end{equation}
and the Pochhammer (or shifted factorial) symbol $(a)_{k}$ is defined by
\begin{equation}
	(a)_{0} = 1\,, \qquad (a)_{k} = a (a+1)(a+2)  \cdots (a+k-1) \,, 
	\quad k = 1, 2, \ldots \,. 
\end{equation}

\subsubsection{Tratnik cases}

For the case $D=3$ with $R_{12}=0$ in Eq. \eqref{Rtratnik}, $Q_{\vec x}(\vec 
k)$ can be expressed as a product of two uni-variate Krawtchouk 
polynomials \cite{GVZ2103} (see also \cite{Iliev2007} and the appendix of \cite{geronimo2010bispectrality}),
\begin{equation}
Q_{x_{1}, x_{2}, x_{3}}(k_{1}, k_{2}, k_{3})=\frac{1}{(-N)_{x_{1}+x_{2}}} 
\mathcal{k}_{x_{1}}(k_{1};p_{1};N-x_{2})\,
\mathcal{k}_{x_{2}}\Big(k_{2};\frac{p_{2}}{1-p_{1}};N-k_{1}\Big) \,,
\label{QTratnik}
\end{equation}
where $\mathcal{k}_{x}(k;p;N)$ is given by \eqref{Krawt}.

For the case $D=4$ with the rotation matrix given by \eqref{Rtratnik4},
the functions $Q_{\vec x}(\vec k)$ can be expressed as a product of three uni-variate Krawtchouk polynomials \cite{Iliev2007}\footnote{To avoid indeterminate values when evaluating these functions numerically, it is convenient to define the uni-variate Krawtchouk polynomials $\mathcal{k}_{x}(k;p;N)$ \eqref{Krawt} to be $0$ if $N < \min(x,k)$.},
\begin{multline}
Q_{\vec x}(\vec k)=\frac{1}{(-N)_{x_{1}+x_{2}+x_3}} 
\mathcal{k}_{x_{1}}(k_{1};p_{1};N-x_{2}-x_{3})\,
\mathcal{k}_{x_{2}}\Big(k_{2};\frac{p_{2}}{1-p_{1}};N-x_{3}-k_{1}\Big) \\ \times \mathcal{k}_{x_{3}}\Big(k_{3};\frac{p_{3}}{1-p_{1}-p_{2}};N-k_{1}-k_{2}\Big)\,.
\label{QTratnikD4}
\end{multline}

\subsubsection{One-parameter cases}\label{sec:oneP}

For the $D=3$ case with one parameter \eqref{Roneparam}, the normalized eigenfunctions are given in terms of a single uni-variate Krawtchouk polynomial \cite{GVZ2103},
\begin{multline}
\langle \vec k |\vec x \rangle = \delta_{k_1, x_1}
\sqrt{\frac{(-1)^{x_2}(N-k_1)!} {k_2! x_2! (N-k_1-k_2)!\, (k_1 - N)_{x_2}}} \\ \times (1-p)^{\frac{1}{2}(N-k_1-k_2-x_2)} p^{\frac{1}{2}(k_2 + x_2)}\, \mathcal{k}_{x_2}(k_2; p; N-k_1) \,.
\label{eigoneparam}
\end{multline}

Similarly, for the $D=4$ case with one parameter \eqref{RoneparamD4}, the normalized eigenfunctions are given by
\begin{multline}
\langle \vec k |\vec x \rangle = \delta_{k_1, x_1}\, \delta_{k_2, x_2}
\sqrt{\frac{(-1)^{x_1+x_2+x_3}N! x_1! x_2!} {k_1! k_2! k_3! x_3! (N-k_1-k_2-k_3)!\, (- N)_{x_1+x_2+x_3}}} \\
 \times (1-p)^{\frac{1}{2}(N-k_1-k_2-k_3-x_3)} p^{\frac{1}{2}(k_3 + x_3)}\, \mathcal{k}_{x_3}(k_3; p; N-k_1-k_2) \,.
\label{eigoneparamD4}
\end{multline}

\subsubsection{A derivation of the overlap coefficients}

The explicit results for overlap coefficients given above, which will be needed for the entanglement entropy computations in Sec. \ref{sec:entanglement}, were borrowed from the literature. For completeness, we sketch here a way of deriving such results, see also \cite{iliev2012lie,mizukawa2004n+}.

Denote by $v_1,\dots,v_D$ the standard orthonormal basis vectors of $\mathbb{C}^D$ and $(E_{ij})_{i,j=1,\dots,D}$ the elementary $D\times D$ matrices, so that $E_{ij}\cdot v_k=\delta_{j,k}v_i$. The matrices $E_{ij}$ act on $(\mathbb{C}^D)^{\otimes N}$ by
\begin{equation}
E_{ij}\cdot v_{i_1}\otimes\dots\otimes v_{i_N}=\sum_{a=1}^N v_{i_1}\otimes\dots\otimes E_{ij}\cdot v_{i_a}\otimes\dots \otimes v_{i_N}\,.
\end{equation}
We will use the following notation for the natural orthonormal basis of $(\mathbb{C}^D)^{\otimes N}$:
\begin{equation}
v_{(i_1,\dots,i_N)}=v_{i_1}\otimes\dots\otimes v_{i_N}\,,\quad\ \ \ \ \ \ \ i_1,\dots,i_N\in\{1,\dots,D\}\,.
\end{equation}
The basis vectors $|\vec{x}\rangle$ in bijection with the lattice sites are realized as an orthonormal basis of the symmetric part in the tensor product $(\mathbb{C}^D)^{\otimes N}$. More precisely, consider the $N$-th symmetrizer, that is the normalized sum over all permutations in $S_N$ of the factors in the tensor product:
\begin{equation}
P\cdot v_{(i_1,\dots,i_N)}=\frac{1}{N!}\sum_{\pi\in S_N} v_{(i_{\pi(1)},\dots,i_{\pi(N)})}\,.
\end{equation}
Note that $P$ commutes with $E_{ij}$. Then we set
\begin{equation}
|\vec{x}\rangle=|x_1,\dots, x_D\rangle=\frac{\sqrt{N!}}{\sqrt{x_1!\dots x_D!}}\,P\cdot v_{(\underbrace{1\,\!,\dots,1}_{x_1},\dots\dots\dots,\underbrace{D\,\!,\dots,D}_{x_D})}\,,\ \ \qquad x_1+\dots+x_D=N \,.
\label{vecxwithP}
\end{equation}
The prefactor ensures that the vectors $|\vec{x}\rangle$ are orthonormal. Note that $|\vec{x}\rangle$ is also obtained by applying~$P$, with the same prefactor, on $v_{(i_1,\dots,i_N)}$ whenever $(i_1,\dots,i_N)$ contains $x_1$ times $1$, $x_2$ times $2$, and so on. 

It is easy to check that the action of $E_{ij}$ on these vectors is
\begin{equation}
E_{ij}|\vec{x}\rangle=a_i^{\dagger}a_j|\vec{x}\rangle, \quad i,j=1,\dots,D\,,
\end{equation}
where $a_i^{\dagger}, a_j$ are as in \eqref{arep}.

Now recall that $R\in SO(D)$ and define $U(R)$ on $(\mathbb{C}^D)^{\otimes N}$ by $R\otimes \dots\otimes R$, that is
\begin{equation}
U(R)\cdot v_{(i_1,\dots,i_N)}=\sum_{a_1,\dots,a_N} R_{a_1i_1}\dots R_{a_Ni_N}v_{(a_1,\dots,a_N)}\,.
\label{actionR}
\end{equation}
The operators $P$ and $U(R)$ commute, so that $U(R)$ restricts to the symmetrized product. Therefore, $U(R)$ acts on $|\vec{x}\rangle$. The operators $X_i$ and $H_i$ acting on $|\vec{x}\rangle$ are defined by
\begin{equation}
X_i=E_{ii}, \qquad  H_i=U(R)^{\dagger}X_iU(R)=\sum_{j,k=1}^DR_{ij}R_{ik}E_{jk}\ ,
\end{equation}
in agreement with \eqref{XH}.
To obtain the eigenvectors $| \vec{k}\rangle$ of $H_i$, we need to apply $U(R)^{\dagger}$ to the eigenbasis $|\vec{x}\rangle$ of $X_i$, and for this we are going to use \eqref{vecxwithP}, the fact that $U(R)$ commutes with $P$ and \eqref{actionR}.
\paragraph{$\boldsymbol{D=2}$.} Let us detail the calculation for $D=2$. We have, for $k_1+k_2=N$,
\begin{align}
|k_1,k_2\rangle& =\frac{\sqrt{N!}}{\sqrt{k_1!k_2!}}U(R)^{\dagger}P\cdot v_{(\underbrace{1\dots,1}_{k_1},\underbrace{2\dots,2}_{k_2})} \non\\
&=\frac{\sqrt{N!}}{\sqrt{k_1!k_2!}}PU(R)^{\dagger}\cdot v_{(\underbrace{1\dots,1}_{k_1},\underbrace{2\dots,2}_{k_2})} \non\\
& = \displaystyle \frac{\sqrt{N!}}{\sqrt{k_1!k_2!}}P \sum_{a_1,\dots,a_{k_1},b_1,\dots,b_{k_2}} R_{1a_1}\dots R_{1a_{k_1}}R_{2b_1}\dots R_{2b_{k_2}}v_{(a_1,\dots,a_{k_1},b_1,\dots,b_{k_2})} \non \\
& = \displaystyle \sum_{x_1,x_2} \frac{\sqrt{x_1!x_2!}}{\sqrt{k_1!k_2!}}\left( \sum_{i=0}^{x_1} \binom{k_1}{i}\binom{k_2}{x_1-i}R_{11}^iR_{21}^{x_1-i}R_{12}^{k_1-i}R_{22}^{k_2-(x_1-i)}\ \right) |x_1,x_2\rangle \,.
\label{D2overlap}
\end{align}
For the last equality, we collect all terms in the sum with $x_1$ occurrences of $1$ in $(a_1,\dots,a_{k_1},b_1,\dots,b_{k_2})$. We can choose $i$ of them among $(a_1,\dots,a_{k_1})$, resulting in $R_{11}^i$, and $(x_1-i)$ among $(b_1,\dots,b_{k_2})$, resulting in $R_{21}^{x_1-i}$.

If we take the matrix
$R=\left(\begin{array}{cc} \sqrt{1-p} & \sqrt{p} \\ -\sqrt{p} & \sqrt{1-p} \end{array}\right)$
as in \eqref{RKraw}, we find
\begin{multline}
\langle x_1,x_2|k_1,k_2\rangle = \frac{\sqrt{x_1!(N-x_1)!}}{\sqrt{k_1!(N-k_1)!}} (\sqrt{1-p})^{N-(k_1+x_1)}(\sqrt{p})^{x_1+k_1}(-1)^{x_1}\\ \times \sum_{i=0}^{x_1} \binom{k_1}{i}\binom{N-k_1}{x_1-i}(-1)^i(1-p)^{i}p^{-i}\,.
\end{multline}
We recover \eqref{overlap} for $D=2$ with \eqref{QKrawt}, see e.g. \cite{MathWorld}.

\paragraph{Arbitrary $\boldsymbol{D}$.}
Define the binomial, and more generally multinomial coefficients, as
\begin{equation}
\binom{k}{a,b}=\frac{k!}{a!b!}\, , \qquad \binom{k}{a_1,\dots,a_D}=\frac{k!}{a_1!\dots a_D!}\,.
\end{equation}
We can write the formulas for the overlap coefficients for $D=2$ in \eqref{D2overlap} as
\begin{equation}
\frac{\sqrt{x_1!x_2!}}{\sqrt{k_1!k_2!}}\sum_{(a_{ij})} \binom{k_1}{a_{11},a_{21}}\binom{k_2}{a_{12},a_{22}}\prod_{\mu,\nu}R_{\nu\mu}^{a_{\mu\nu}}\,,
\end{equation}
where the sum is over nonnegative integers $a_{11},a_{12},a_{21},a_{22}$ satisfying:
\[\begin{array}{cccl}
a_{11} & a_{12} & \rightarrow & x_1 \qquad\qquad\text{(the sum of this line is $x_1$)}\\
a_{21} & a_{22} & \rightarrow  & x_2 \qquad\qquad\text{(the sum of this line is $x_2$)}\\[0.5em]
\downarrow & \downarrow \\
k_1 & k_2 & & \qquad\qquad\text{(the sums of the columns are respectively $k_1$ and $k_2$).}
\end{array}\]
Of course, here there is only one independent index $a_{11}$, which can go from $0$ to $x_1$. Reproducing straightforwardly the reasoning made for $D=2$, one obtains the following result for general $D$,
\begin{equation}
\langle x_1,\dots,x_D|k_1,\dots ,k_D\rangle = \displaystyle \frac{\sqrt{x_1! \ldots x_D!}}{\sqrt{k_1! \ldots k_D!}} \sum_{(a_{ij})} \binom{k_1}{a_{11},\dots,a_{D1}}\dots \binom{k_D}{a_{1D},\dots,a_{DD}}\prod_{\mu,\nu}R_{\nu\mu}^{a_{\mu\nu}}\,,
\end{equation}
where the sum is now over nonnegative integers $(a_{ij})_{i,j=1,\dots ,D}$ such that the sum of the $r$-th line gives $x_r$ and the sum of the $r$-th column is $k_r$.

\section{Correlation matrix and algebraic Heun operator}\label{sec:cormat}

In this section, we define the correlation matrix and express its entries in terms of overlaps, which are given in Sec. \ref{sec:eigenfunctions}. We then construct, for any $D$, an operator that commutes with the chopped correlation matrix, thereby generalizing the $D=2$
tridiagonal Heun operator found in \cite{Crampe:2019upj, Crampe:2021}.

\subsection{Correlation matrix}

Let us consider a fermionic eigenstate $|\Psi\rangle\!\rangle$ as 
in \eqref{Fermioneigenstates}, where the set $\setF\subseteq V$ is given by 
\begin{equation}
\setF = \{ \vec k \in V | \vec\alpha \cdot \vec k \le K \} \,, \qquad 0 
\le K \le N \,.
\label{tildeSset}
\end{equation}
The correlation matrix in this state, which will be needed later for the 
entanglement entropy computation, is given by the matrix elements
\begin{equation}
\hat C_{\vec x', \vec x} = \langle\! \langle \Psi | 
c^{\dagger}_{\vec x'}\, c_{\vec x}\, |\Psi\rangle\!\rangle
= \sum_{\vec k', \vec k \in V} 
\langle \vec x' | \vec k' \rangle
\langle \vec x |  \vec k \rangle
\langle\! \langle \Psi | \tilde{c}^{\dagger}_{\vec k'}\, 
\tilde{c}_{\vec k}\, |\Psi\rangle\!\rangle \,.
\label{cormat1}
\end{equation}
We find  using \eqref{ctildeccr} and \eqref{Fermioneigenstates} that
\begin{equation}
\langle\! \langle \Psi | \tilde{c}^{\dagger}_{\vec k'}\, \tilde{c}_{\vec k}\, |\Psi\rangle\!\rangle 
= \begin{cases}
1 & \text{  if  } \quad\vec k = \vec k' \in \setF \\
0 & \text{  otherwise  } 
\end{cases} \,.
\end{equation}
The correlation matrix elements \eqref{cormat1} are therefore given by
\begin{equation}
\hat C_{\vec x', \vec x} = \sum_{\vec k \in \setF} 
\langle \vec x' |  \vec k \rangle
\langle \vec k | \vec x\rangle\,,
\label{cormat2}
\end{equation}
which implies that the correlation matrix is
\begin{equation}
\hat C = \sum_{\vec x', \vec x \in V} |\vec x'\rangle \hat C_{\vec 
x', \vec x}\, \langle \vec x | = \sum_{\vec k \in \setF} 
|  \vec k \rangle \langle \vec k | \,.
\end{equation}

\subsection{Algebraic Heun operator}\label{sec:Heun}

\begin{figure}[h]
    \centering
\begin{tikzpicture}
\node[inner sep=0pt] (russell) at (0,0) {\includegraphics[scale = 0.7]{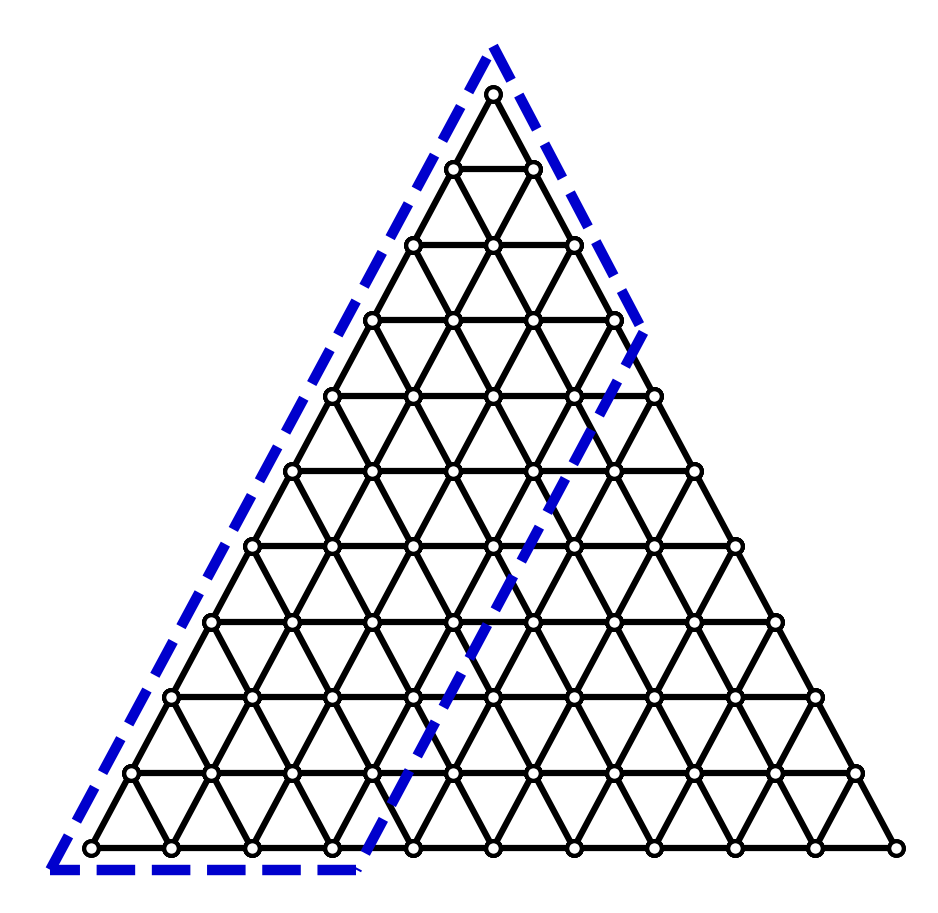}};
\node at (-2,0) {\color{blue}\Large $\mathcal{A}$};
\end{tikzpicture}
    \caption{Example of subsystem $\mathcal{A}$ for $D=3$ when $\vec \beta$ is a unit vector, $N=10$ and $L=3$.}
    \label{fig:subsyst}
\end{figure}

Recall that the set $\setF$ is defined in \eqref{tildeSset}, and 
let us now define a similar set $\setA \subseteq V$ by
\begin{equation}
\setA = \{ \vec x \in V | \vec\beta \cdot \vec x \le L \} \,, \qquad 0 
\le L \le N \,,
\label{Sset}
\end{equation}
where $\vec \beta = (\beta_{1}, \ldots, \beta_{D})$ with $\beta_i \in \{0, 
1 \}$. If $D=2$ and $\vec \beta$ is a unit vector, $\setA=\{0,\dots,L\}$ is a segment of length $L+1$. For $D=3$ and if $\vec \beta$ is a unit vector, subsystem $\setA$ is the triangular lattice from which a triangular part has been removed, see Fig. \ref{fig:subsyst}.

The projection operators on subsets $\setA$ and $\setF$ are
\begin{equation}
\pi_{1} = \sum_{\vec x \in \setA} |\vec x\rangle \langle \vec x | \,,  \qquad
\pi_{2} = \sum_{\vec k \in \setF} |\vec k\rangle \langle \vec k | \, ,
\end{equation}
respectively. We consider a general symmetric bilinear expression of the bispectral operators
$X_{i}$ and $H_{i}$ introduced in Sec. \ref{sec:Solution},
\begin{equation}
	T = \sum_{i=1}^{D} \mu_{i} X_{i} + \sum_{i=1}^{D} \nu_{i} H_{i} + 
	\sum_{i,j=1}^{D} \rho_{ij} \{H_{i} \,, X_{j} \} \,.
\label{Heun}
\end{equation}
This provides an extension to higher dimensional bispectral problems of the algebraic Heun operator introduced in \cite{grunbaum2018algebraic}, see also \cite{eisler2013free, eisler2018properties, Crampe:2019upj, Crampe:2021}. We solve for the coefficients $\mu_{i}, \nu_{i},  \rho_{ij}$ such that $T$ commutes with both $\pi_{1}$ and $\pi_{2}$.

Let us first consider the commutativity of $T$ with $\pi_{1}$. We observe that 
$\vec x \in \setA$ iff $\sum_{i \in \mathcal{b}} x_{i} \le L$,
for some set $\mathcal{b} \subseteq \{1, \ldots, D \}$ 
corresponding to the choice of $\vec \beta$. Furthermore,
$\left[T\,,\pi_{1} \right] = 0$ iff $\langle \vec x' | T | \vec x 
\rangle = 0$ for all $\vec x$ and $\vec x'$ such that
\begin{equation}
	\sum_{i \in \mathcal{b}} x_{i} \le L\,, \qquad \sum_{i \in 
	\mathcal{b}} x'_{i} > L \,.
\label{xy1}
\end{equation}
The fact that $T$ is linear in $H_{i}$ implies that $T$ is linear in $E_{jk}:= 
a^{\dagger}_{j} a_{k}$ with $j \ne k$, which has the property $E_{jk} 
|\vec x \rangle = |\vec x -\vec \epsilon_{j} + \vec \epsilon_{k} 
\rangle$. Therefore, $\langle \vec x' | T | \vec x 
\rangle = 0$ unless
\begin{equation}
	\sum_{i \in \mathcal{b}} x_{i} = L\,, \qquad \sum_{i \in 
	\mathcal{b}} x'_{i} = L + 1 \,.
\label{xy2}
\end{equation}
Let us consider $\vec x$ and $\vec x'$ as in \eqref{xy2}. Then
\begin{equation}
\langle \vec x' | T | \vec x \rangle = \sum_{i=1}^{D} \Bigl[\nu_{i} + 
\sum_{j=1}^{D} \rho_{ij} (x'_{j} + x_{j}) \Bigr] 
\langle \vec x' | H_{i} | \vec x \rangle \,.
\label{step1}
\end{equation}
We choose 
\begin{equation}
\rho_{ij} = 
\begin{cases}
0 & \text{  if  } \quad j \notin \mathcal{b} \\
1 & \text{  if  } \quad j \in \mathcal{b} 
\end{cases} \,.
\end{equation}
It then follows from \eqref{step1} that
\begin{equation}
\langle \vec x' | T | \vec x \rangle = \sum_{i=1}^{D} \Bigl[\nu_{i} + 
\sum_{j \in \mathcal{b}}  (x'_{j} + x_{j}) \Bigr] 
\langle \vec x' | H_{i} | \vec x \rangle \non\\
= \sum_{i=1}^{D} \Bigl[\nu_{i} + (2L+1) \Bigr] \langle \vec x' | H_{i} 
| \vec x \rangle\,,
\end{equation}
which vanishes for $\nu_{i} = -(2L +1)$. 

For the commutativity of $T$ with $\pi_{2}$, we repeat the 
calculation in the $|\vec k\rangle$ basis. In particular, 
$\vec k \in \setF$ iff $\sum_{i \in \mathcal{f}} k_{i} 
\le K$, where $\mathcal{f} \subseteq \{1, \ldots, D \}$
corresponds to the choice of $\vec \alpha$.
The fact that $T$ is linear in $X_{i}$ implies that
$\langle \vec k' | T | \vec k \rangle = 0$ unless   
\begin{equation}
	\sum_{i \in \mathcal{f}} k_{i} = K\,, \qquad \sum_{i \in 
	\mathcal{f}} k'_{i} = K + 1 \,.
\label{kkp}
\end{equation}
Setting
\begin{equation}
\rho_{ij} = 
\begin{cases}
0 & \text{  if  } \quad i \notin \mathcal{f} \\
1 & \text{  if  } \quad i \in \mathcal{f} 
\end{cases} \,,
\end{equation}
we obtain $\mu_{j} = -(2K +1)$. 

In conclusion, the operator $T$ in \eqref{Heun} commutes with both projectors,
\begin{equation}
\left[T\,,\pi_{1} \right] = 0\,, \qquad  \left[T\,,\pi_{2} 
\right] = 0 
\end{equation}
for the coefficient values
\begin{equation}
\mu_{i} = -(2K +1) \beta_{i}\,, \qquad \nu_{i} = -(2L +1) 
\alpha_{i}\,, \qquad  \rho_{ij}  = \alpha_{i} \beta_{j} \,.
\label{eq:coeffHeun}
\end{equation}	
It follows that 
\begin{equation}
	\left[T\,, C \right] = 0\,,
\end{equation}
where $C$ is the chopped correlation matrix
\begin{equation}
    C = \pi_{1} \pi_{2} \pi_{1} \,.
    \label{choppedC}
\end{equation}
We have checked numerically for various examples that the eigenvalues (apart from $0$) of the chopped Heun operator $\pi_1\, T\, \pi_1 = \pi_1\, T$ are not degenerate.
For $D=2$, the result in \cite{Crampe:2019upj, Crampe:2021} for the Heun operator is recovered.

While the eigenvalues of $C$ are expected to agglomerate near $0$ and $1$, the spectrum of $T$ is usually well spaced and thus free of such problem. Since both matrices share a common eigenbasis, the Heun operator $T$ offers an interesting tool to diagonalize the chopped correlation matrix $C$. It also opens the door to the application of Bethe ansatz methods, which was done for the case $D = 2$ \cite{bernard2021heun}. For these reasons, we highlight Eqs. \eqref{Heun} and \eqref{eq:coeffHeun} as a main result of this paper.

\section{Entanglement entropy}\label{sec:entanglement}

In this section, we investigate the entanglement entropy of certain eigenstates $|\Psi\rrangle$ of the Hamiltonian~\eqref{Hamiltonian}. Because the Hamiltonian is quadratic in terms of the fermionic operators, the reduced density matrix associated to a region $\setA$ is a Gaussian operator whose eigenvalues can be obtained \cite{CP01, peschel2003calculation, PE09} from the chopped correlation matrix $C$ in \eqref{choppedC}. Accordingly, the entanglement entropy \eqref{EE} reads
\begin{equation}
\label{eq:SvNCorr}
    S_{vN} = -\Tr( C \log C +(\id-C)\log(\id-C)).
\end{equation}

In studying the entanglement entropy as a function of $N$ for given values of $D$, $\vec\alpha$, and $\vec\beta$, it is also necessary to specify $K$ and $L$ which define the sets $\setF$ in \eqref{tildeSset} and $\setA$ in \eqref{Sset}, respectively.

Let us note that if $\vec \alpha = (1, 1,\dots, 1)$, then the entanglement entropy trivially vanishes. Indeed, in that case we have $\vec \alpha \cdot \vec k = \sum_{i=1}^D k_i=N$ for all $\vec k \in V$. From the definition \eqref{tildeSset} of $\setF$, it follows that for this case $|\setF|=0$ if $K<N$ and $|\setF|=n(N,D)$ if $K=N$, where $n(N,D)$ is given by \eqref{dim}. These situations correspond to a filling fraction $|\setF|/n(N,D)$ equal to
$0$ and $1$, respectively, and in both cases the system is in a product state with zero entanglement entropy.

A natural choice is to consider $K$ such that the system is at half filling, namely that the number of elements in $\setF$ is $|\setF| = \frac{1}{2}n(N,D)$. Moreover, for simplicity, we always choose $\vec \alpha$ such that $\vec\alpha=\vec\epsilon_i$ for some value of $i$ (see Eq. \eqref{epsi}).  In that case, the choice of $K$ does not depend on $i$. 

For $D=2$ (the one-dimensional chain), the notion of half filling is straightforward, and we choose $K=N/2$. Looking at the definition of $\setF$ in \eqref{tildeSset}, this implies that $|\setF|=\lfloor N/2 \rfloor +1$ (remember that the chain has $N+1$ sites for $D=2$). Hence, we have a perfect half filling for odd values of $N$, whereas for even $N$ the filling is $\frac{N+2}{2(N+1)}$ and tends to half filling in the limit $N \to \infty$. 

For $D>2$ however, it is not always possible to choose a value of $K$ such that we are perfectly at half filling. In these cases, we choose the largest $K$ such that $|\setF|$ is less than or equal to $\frac{1}{2}n(N,D)$. For $D=3,4$, we find that the value of $K$ that satisfies this property is $K=N/D-1$. The formula becomes more involved for larger values of $D$, but we do not consider these cases. 

Similarly, we take $\vec\beta=\vec\epsilon_j$ for some value of $j$, and systematically set $L=K$, for simplicity.

\subsection[The case $D=2$]{The case $\boldsymbol{D=2}$}\label{sec:D2}

For $D=2$, the system is a one-dimensional chain of length $N+1$, and we fix $\vec \alpha=\vec \beta=\vec \epsilon_1$. We note that the entropy does not change if we choose $\vec \alpha=\vec \beta=\vec \epsilon_2$ instead. The entropy also depends on the parameter $p$ (see Eq. \eqref{HD2}), and if instead of $\vec \alpha=\vec \beta$ we had $\vec \alpha \neq \vec \beta$, this would be equivalent to substituting $p$ with $1-p$ in our results. Since we investigate the dependence of the entanglement entropy on $p$, our choice for $\vec \alpha$ and $\vec \beta$ is thus generic. 

\begin{figure}
	\centering
	\includegraphics[width=0.43\hsize]{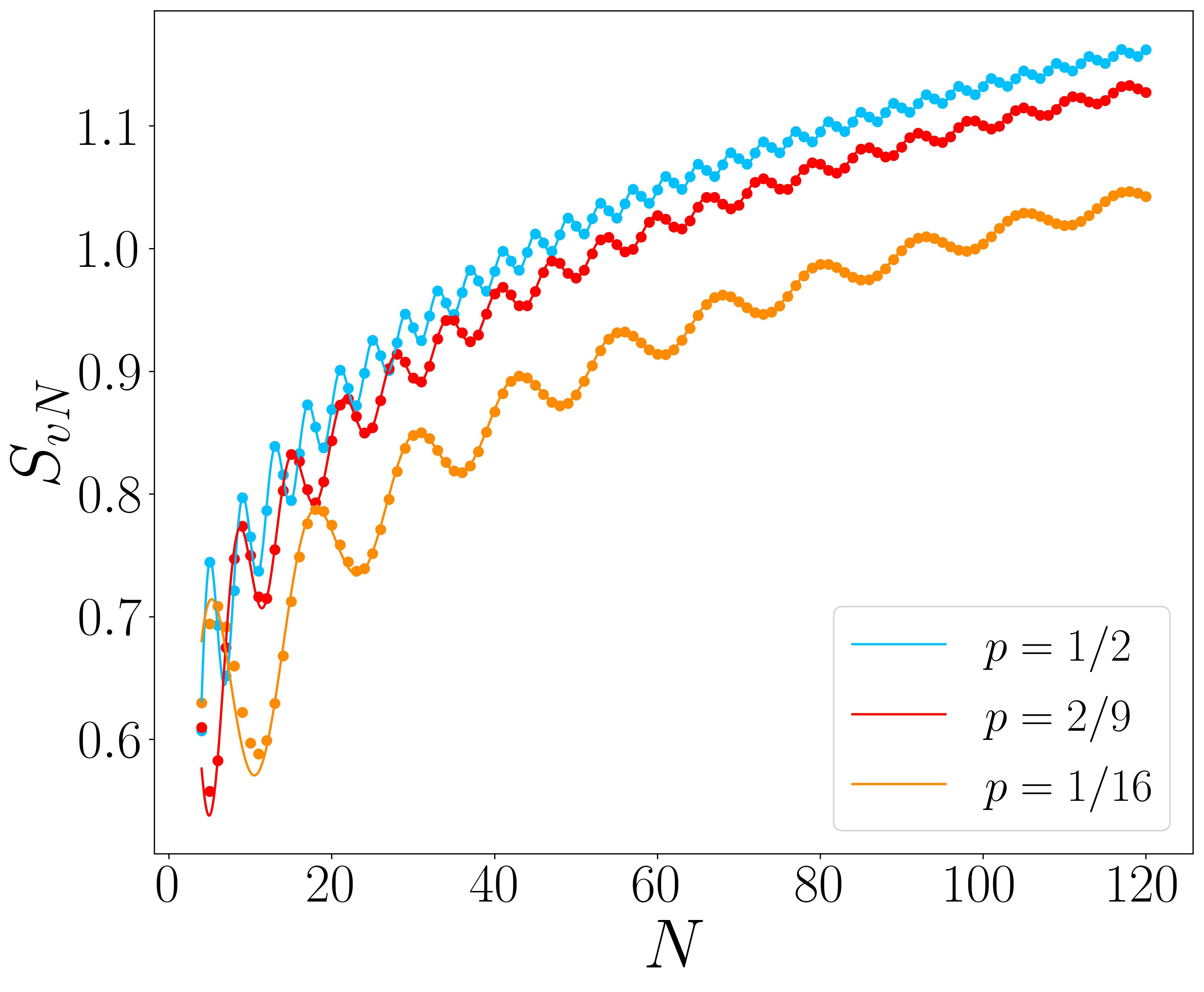}
	\includegraphics[width=0.43\hsize]{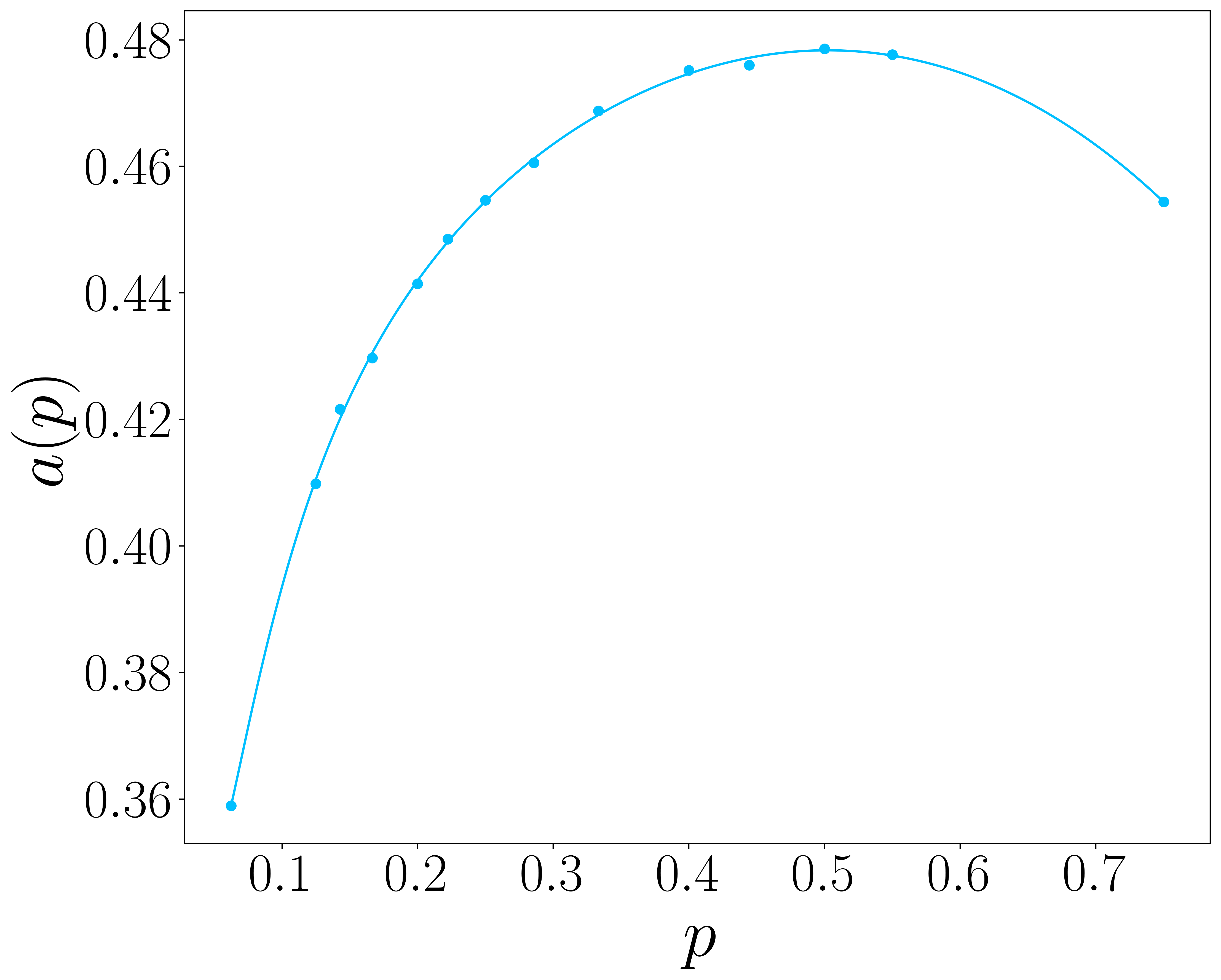}
	\caption{\textit{Left}: Entanglement entropy as a function of $N$ for $D=2$, $\vec \alpha=\vec \beta=\vec \epsilon_1$ and $ K=L=N/2$ for various values of $p$. The sold lines are our conjecture of Eq. \eqref{eq:Svn1dConj}, whereas the dots are obtained from exact diagonalization of the chopped correlation matrix. \textit{Right}: Value of the constant parameter $a(p)$ in Eq. \eqref{eq:Svn1dConj} as a function of $p$. The dots are obtained from numerical fits, and the solid line serves as a guide to the eye.}
	\label{fig:Svn1d}
\end{figure}

We compute the entanglement entropy for $L=K=N/2$ as a function of $N$ for various values of~$p$. Numerically, we use the eigenfunctions given in \eqref{QKrawt} and \eqref{Krawt} to compute the chopped correlation matrix and use Eq. \eqref{eq:SvNCorr}. We conjecture the following result for the entanglement entropy,
\begin{equation}
\label{eq:Svn1dConj}
S_{vN} = \frac 16 \log \left(\frac {N+1}2\right)+ a(p) - \frac{1}{2(N+1)}\frac{\cos(\frac{\pi}{2}\frac{N+1}{m(p)})}{\sin(\frac{\pi}{2m(p)})}+\dots \, ,
\end{equation}
where
\begin{equation}
\label{eq:mp}
    m(p)=\frac{1}{2}\left(1-\log p +\frac{1-\log 2}{2p}\right),
\end{equation}
$a(p)$ is a non-universal constant with respect to $N$, and the ellipsis indicate terms of order smaller than~$N^{-1}$ in the large-$N$ limit. We compare our conjecture with numerical results in the left panel of Fig. \ref{fig:Svn1d} and find excellent agreement, already for moderate values of $N$. We computed the entanglement entropy for numerous values of $p$, and systematically found a similar match between the numerical data and the conjecture of Eq. \eqref{eq:Svn1dConj}. However, we did not include all the corresponding graphs in Fig. \ref{fig:Svn1d} for clarity. 

The leading term $\frac 16 \log \left(\frac {N+1}2\right)$ in Eq. \eqref{eq:Svn1dConj} is in agreement with previous results from Ref. \cite{FA21}. In that paper, the authors used methods from CFT in a curved background \cite{DSVC17} to extract the leading contribution for the special case $p=1/2$ at half filling. In particular, the system is described by a CFT of massless Dirac fermions in a curved $(1+1)$-dimensional background. We find that this leading contribution holds for any values of $p$ at half filling. 

The constant term with respect to $N$ in Eq. \eqref{eq:Svn1dConj}, that we denote $a(p)$, depends non-trivially on the parameter $p$. For $p=1/2$, we find $a(1/2)=0.4786$, which corresponds to the exact results for the homogeneous XX chain \cite{fagotti2011universal} at half filling. This was already observed in Ref. \cite{FA21}. We report our results for $a(p)$ in the right panel of Fig. \ref{fig:Svn1d}. The dots are obtained from numerical fits of our results for the entropies, and the solid line is a guide to the eye. The latter does not reflect an analytical prediction or a conjecture, and therefore we do not report its expression here.

Finally, we observe oscillations in the sub-leading contribution of order $N^{-1}$ in Eq. \eqref{eq:Svn1dConj}. Similar oscillations have already been studied extensively for homogeneous chains \cite{calabrese2010parity,calabrese2010universal,fagotti2011universal,bonsignori2019symmetry,murciano2020symmetry,berthiere2021entanglement} and in the context of the so-called Rainbow chain \cite{rodriguez2017more}. For homogeneous chains, the frequency of the oscillations depends only on the filling fraction. Here, for a fixed filling fraction (we are always at half filling), we observe that the frequency of the oscillations depends on the model through the parameter $p$ and the function $m(p)$ in Eq. \eqref{eq:mp}. As a particular case, for $p=1/2$ we have $m(1/2)=1$ and we recover the results of Ref. \cite{FA21}. The non-trivial dependence of the frequencies of sub-leading oscillations on the parameters of the model is one of our main results, and it calls for further investigations at different filling fractions and for different inhomogeneous chains. These issues will be addressed in a forthcoming publication.

\subsection[The case $D=3$]{The case $\boldsymbol{D=3}$}
In this subsection, we investigate the entanglement entropy for $D=3$, where the whole system is a planar triangle, as depicted on the right of Fig. \ref{fig:D23}. As discussed as the beginning of Sec. \ref{sec:entanglement}, we systematically take $L=K=N/3-1$. 

\subsubsection[Numerical results for Tratnik with $D=3$]{Numerical results for Tratnik with $\boldsymbol{D=3}$}\label{sec:D3Trat}

\begin{figure}
	\centering
	\includegraphics[width=0.43\hsize]{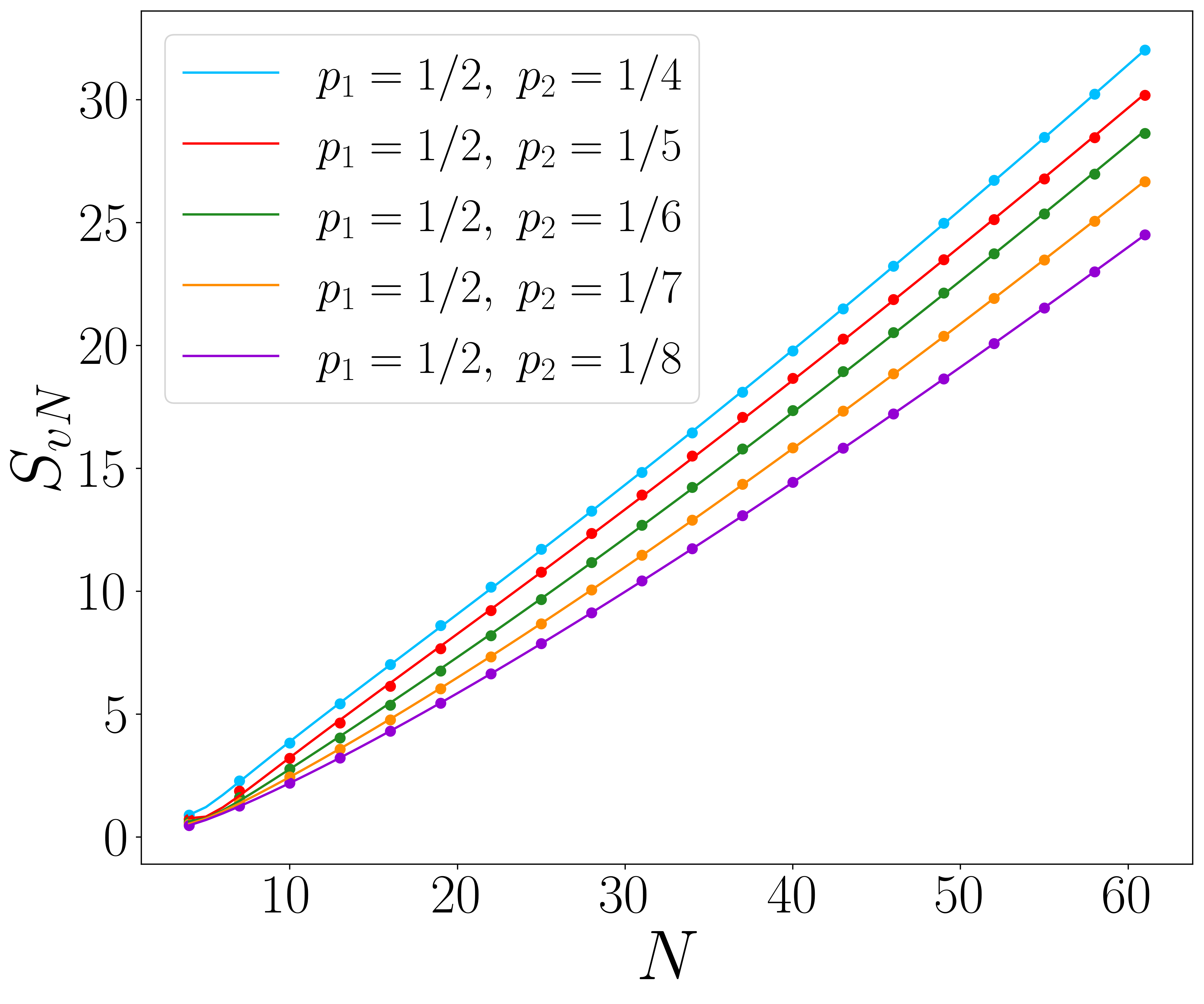}
	\includegraphics[width=0.43\hsize]{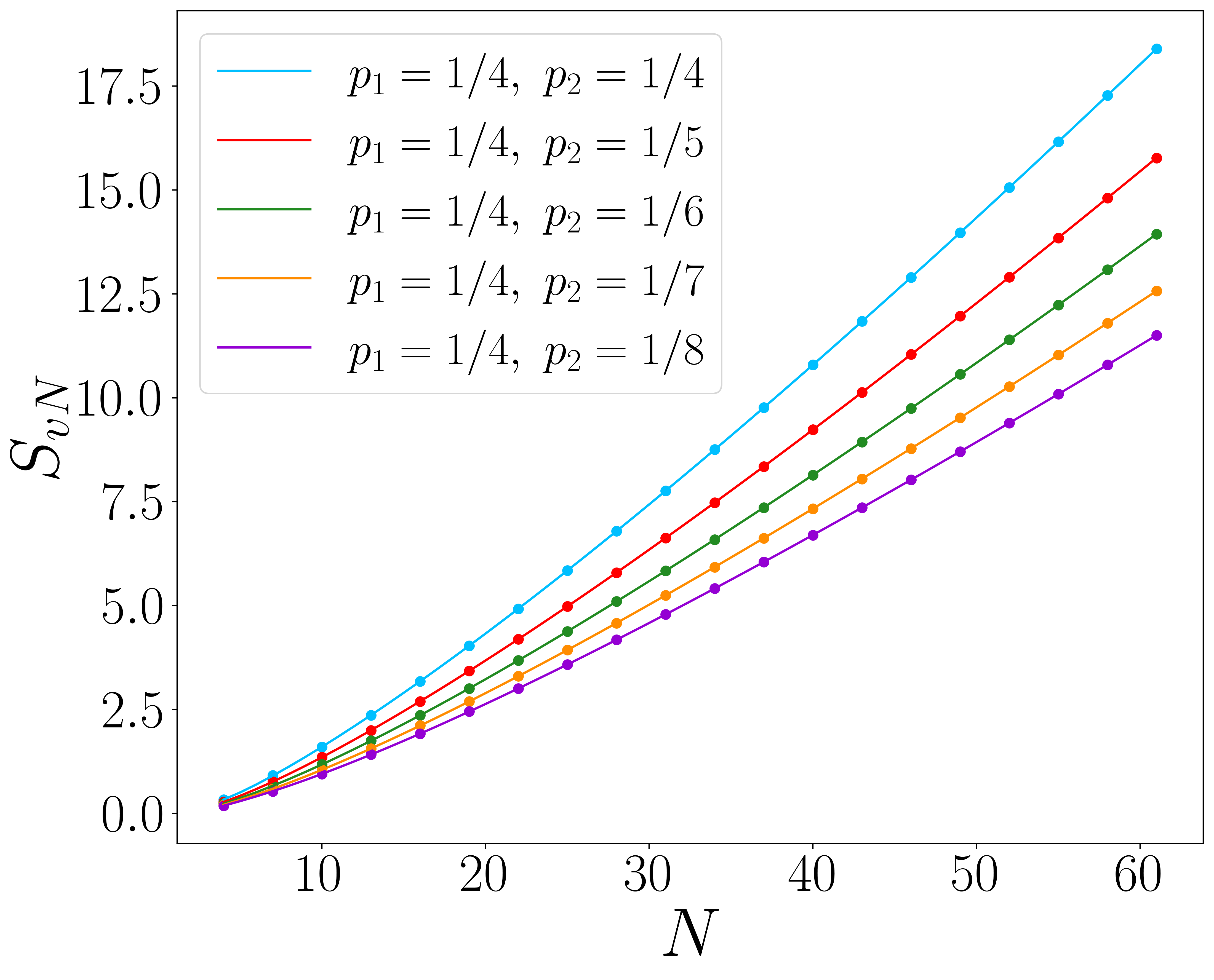} \\
	\includegraphics[width=0.43\hsize]{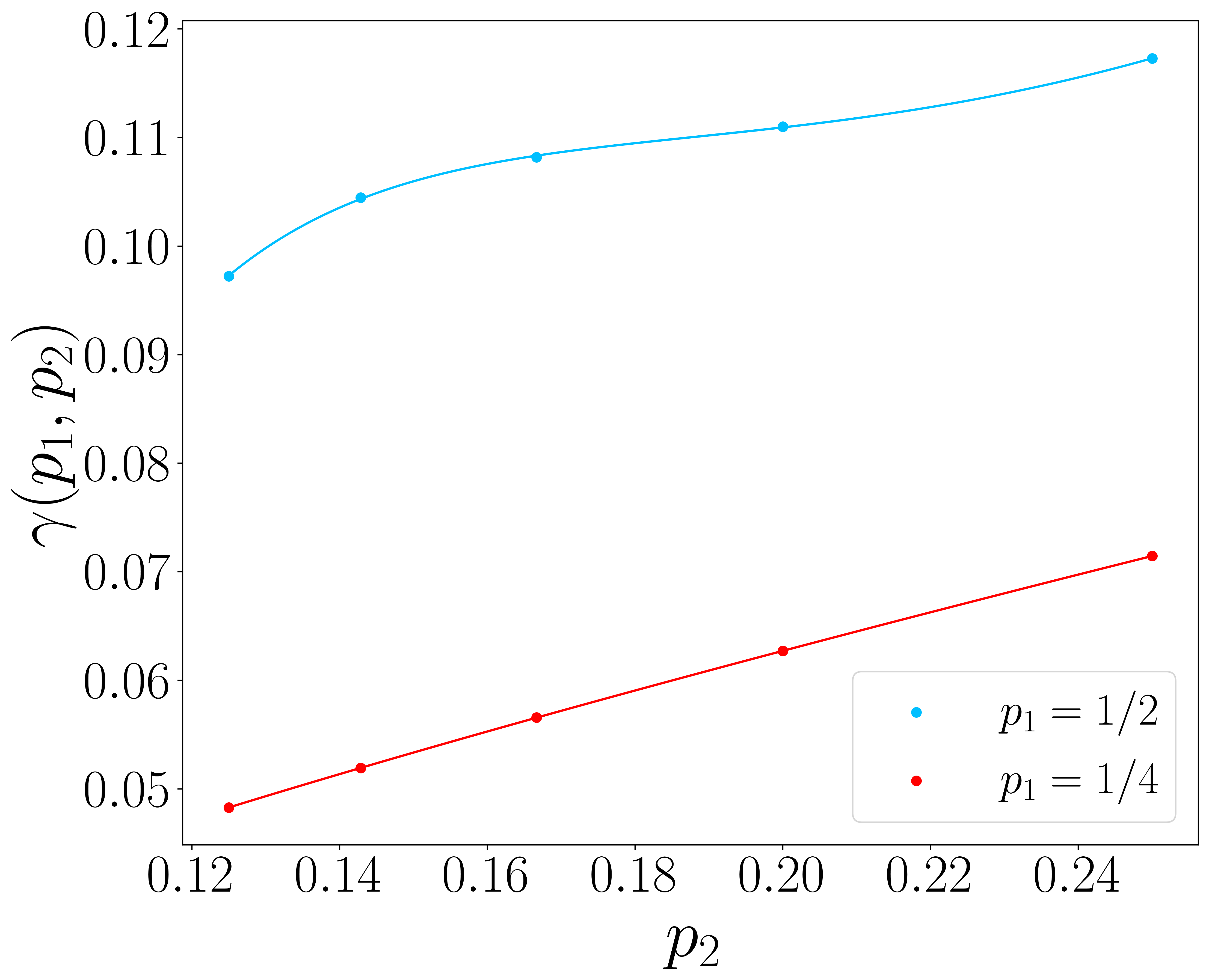}
		\includegraphics[width=0.43\hsize]{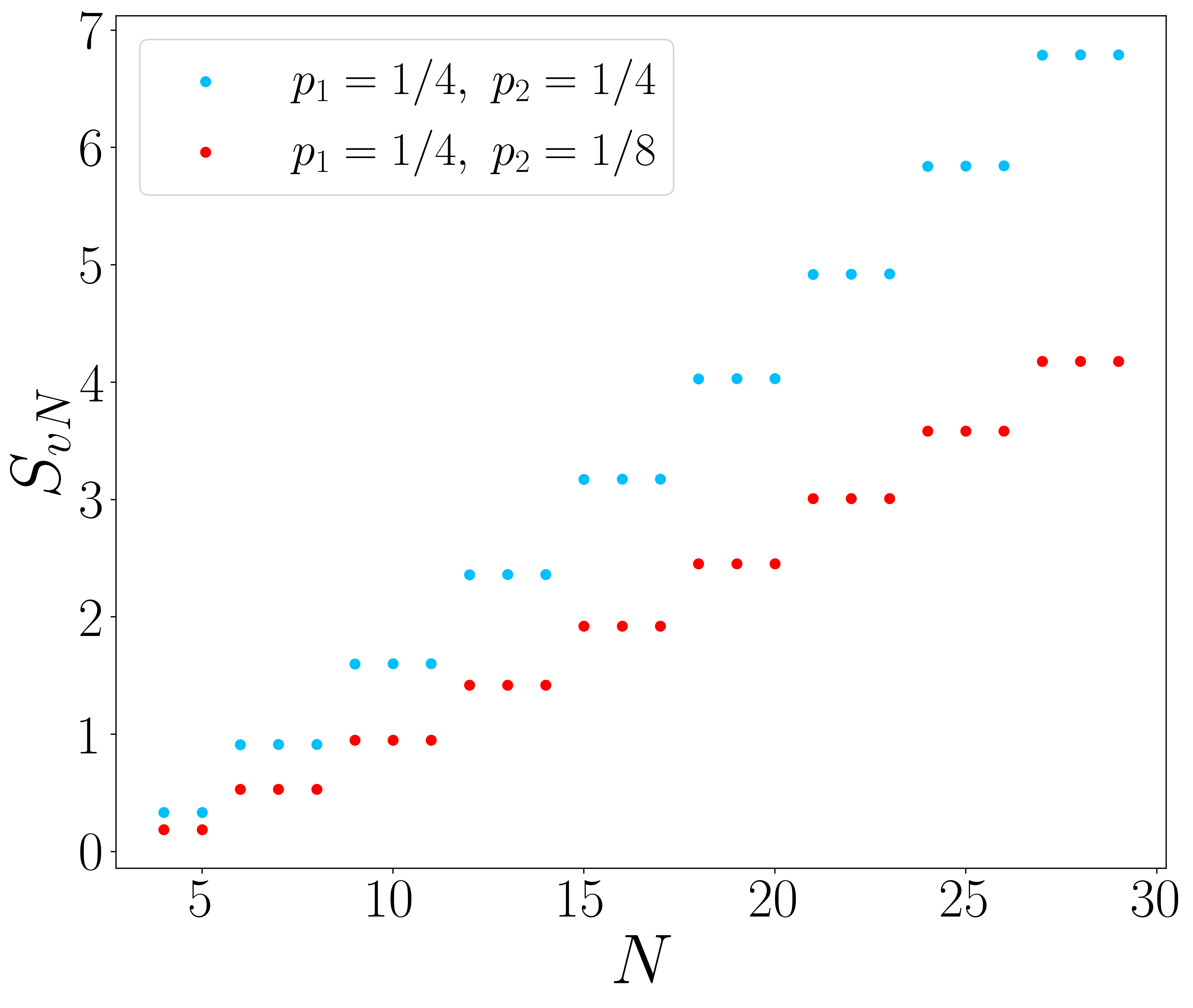}
	
	\caption{\textit{Top}: Entanglement entropy as a function of $N=3n+1$ for the Tratnik case with $D=3$, $L=K=N/3-1$, $\vec{\alpha}=\vec \epsilon_2$, $\vec{\beta}=\vec \epsilon_1$ and various values of $p_1,p_2$. The dots are obtained from direct diagonalization of the chopped correlation matrix, and the solid lines are fits of the form of Eq. \eqref{eq:SvnD3Trat}. \textit{Bottom left}: Leading coefficient $\gamma$ in the expansion of Eq. \eqref{eq:SvnD3Trat} as a function of $p_2$ with $p_1=1/2$ and $p_1=1/4$. The dots are obtained from our numerical data, and the solid lines serve as guides to the eye. \textit{Bottom right}: Entanglement entropy as a function of $N$ for the Tratnik case with $D=3$ and the same parameters as in the top panels. We do not impose $N=3n+1$ and observe clear plateaus of length 3. }
	\label{fig:SvnTratnik}
\end{figure}

We start with the Tratnik case $D=3$ and $R_{12}=0$, where the eigenfunctions $\langle \vec x |\vec k \rangle$ have a simple closed form, see Eqs. \eqref{overlap}, \eqref{W} and \eqref{QTratnik}. The entropy depends on $p_1,p_2$, as well as the system size, the eigenstate under consideration and the subsystem, through the respective choices of $N$, $K$ and $\vec{\alpha}$, and $L$ and $\vec{\beta}$. We compute the entanglement entropy numerically for various values of $p_1$ and $p_2$ with $N=3n+1$, $L=K=N/3-1$, $\vec{\alpha}=\vec \epsilon_2$, and $\vec{\beta}=\vec \epsilon_1$. We display our results on the top panels of Fig. \ref{fig:SvnTratnik}. The dots are the numerical data obtained by direct diagonalization of the chopped correlation matrix, and the solid lines correspond to fits of the form 
\begin{equation}
\label{eq:SvnD3Trat}
    S_{vN} =  \gamma \ N \log N + \dots
\end{equation}
where the ellipsis indicates terms that are sub-leading compared to $N \log N$ in the large-$N$ limit.

For critical free fermions in $d$ dimensions, the entanglement entropy exhibits a logarithmic violation of the area law \cite{ECP08} and scales as $S_{vN}~\sim~N^{d-1}\log N$, where $N$ is a characteristic length scale of the subsystem along one spatial dimension \cite{wolf2006violation,gioev2006entanglement,li2006scaling}. In our case, the fact that the leading contribution to the entanglement entropy in Eq. \eqref{eq:SvnD3Trat} scales as $N \log N$ thus indicates that, similarly to the one-dimensional case discussed in Sec. \ref{sec:D2}, our inhomogeneous system has a similar behavior as critical free fermions in two dimensions. However, here the leading coefficient depends on the parameters $p_1$ and $p_2$ in a non-trivial way, and we restrict ourselves to numerical evaluation of $\gamma$. In the bottom left panel of Fig. \ref{fig:SvnTratnik}, we show the dependence of this coefficient on $p_2$ for fixed values of $p_1$. The dots are obtained from our numerical data, and the solid lines are fits that serve as guides to the eyes. We do not report their expressions.

Another difference with the one-dimensional chain is that here we do not have parameter-dependent oscillations. In the bottom right panel of Fig. \ref{fig:SvnTratnik}, we plot the entanglement entropy with the same parameters as in the top panels, but for all values of $N$ (instead of focusing on $N=3n+1$). We observe plateaus where the entropy has almost the same value for consecutive values of $N$. This phenomenon is very different in nature from the oscillations we observe in the one-dimensional chain, since here the plateaus do not depend on $p_1,p_2$, but instead appear because of our choices for $L$ and $K$. In fact, we verified that for $L=K=N/M-1$ (hence away from half filling for $M \neq 3$), we have plateaus of length $M$. Here they have length $3$ because $L=K=N/3-1$. These plateaus are thus artifacts of the geometry of the model, and we remove them by considering $N=3n+1$.

\subsubsection[Symmetric point for Tratnik at $p_1=1/2$, $p_2=1/4$]{Symmetric point for Tratnik at $\boldsymbol{p_1=1/2}$, $\boldsymbol{p_2=1/4}$}

For the $D=3$ Tratnik case with parameter values $p_1=1/2$, $p_2=1/4$, the nonzero matrix elements of the rotation matrix \eqref{Rtratnik} are $\pm 1/2\,, \pm 1/\sqrt{2}$, and both Krawtchouk polynomials in \eqref{QTratnik} have the same parameter value (namely, $1/2$). For this case, we find numerically that
the entanglement entropies corresponding to various pairs $( \vec\alpha, \vec\beta)$ are equal. For example, all the pairs $\{ (\vec\epsilon_1, \vec\epsilon_1)\,, ( \vec\epsilon_2, \vec\epsilon_2)\,, ( \vec\epsilon_3, \vec\epsilon_2)\,, (\vec\epsilon_1, \vec\epsilon_3) \}$ correspond to the same entanglement entropy, and similarly for $\{ (\vec\epsilon_2, \vec\epsilon_1)\,, ( \vec\epsilon_3, \vec\epsilon_1)\,, ( \vec\epsilon_2, \vec\epsilon_3)\,, (\vec\epsilon_3, \vec\epsilon_3) \}$.

Many of the above equivalences in entanglement entropy can be explained by the existence of simple changes of basis for this special case. Indeed, consider the transformation ${\cal P}^{(1)}$ defined by
\begin{equation}
    {\cal P}^{(1)} |x_1, x_2, x_3 \rangle =  |x_3, x_2, x_1 \rangle \,,
\end{equation}
whose action on the $k$-basis is given by
\begin{equation}
    {\cal P}^{(1)} |k_1, k_2, k_3 \rangle =  (-1)^{N+k_1}|k_1, k_3, k_2 \rangle \,.
\end{equation}
Conjugation by ${\cal P}^{(1)}$ therefore gives
\begin{equation}{\cal P}^{(1)}\ :\ 
\left\{\begin{array}{l}
X_1\mapsto X_3\\[0.2em]
X_2\mapsto X_2\\[0.2em]
X_3\mapsto X_1
\end{array}\right.\ \ \ \ \text{and}\ \ \ \ \ \left\{\begin{array}{l}
H_1\mapsto H_1\\[0.2em]
H_2\mapsto H_3\\[0.2em]
H_3\mapsto H_2
\end{array}\right. \,.
\end{equation}
Similarly, conjugation by ${\cal P}^{(1)}$ on the projectors gives
\begin{equation}{\cal P}^{(1)}\ :\ 
\left\{\begin{array}{l}
\pi_{X_1}\mapsto \pi_{X_3}\\[0.2em]
\pi_{X_2}\mapsto \pi_{X_2}\\[0.2em]
\pi_{X_3}\mapsto \pi_{X_1}
\end{array}\right.\ \ \ \ \text{and}\ \ \ \ \ \left\{\begin{array}{l}
\pi_{H_1}\mapsto \pi_{H_1}\\[0.2em]
\pi_{H_2}\mapsto \pi_{H_3}\\[0.2em]
\pi_{H_3}\mapsto \pi_{H_2}
\end{array}\right. \,,
\end{equation}
where $\pi_{X_i}$ and $\pi_{H_i}$ 
denote the projectors $\pi_1$ with $\beta=\epsilon_i$, 
and $\pi_2$ with $\alpha=\epsilon_i$, respectively.
Consequently, the chopped correlation matrices 
$\pi_{X_1}\pi_{H_1}\pi_{X_1}$ and
$\pi_{X_3}\pi_{H_1}\pi_{X_3}$ are related by conjugation, which explains the fact that $(\vec\epsilon_1, \vec\epsilon_1)$ and
$(\vec\epsilon_1, \vec\epsilon_3)$ have the same entanglement entropy. With this argument, we find the following equivalences:
\begin{equation}\label{equivalence}
\begin{array}{c}
(\vec{\epsilon_1},\vec{\epsilon_1})\sim (\vec{\epsilon_1},\vec{\epsilon_3})\,, \\[0.4em]
(\vec{\epsilon_2},\vec{\epsilon_2})\sim (\vec{\epsilon_3},\vec{\epsilon_2})\,, \\[0.4em]
(\vec{\epsilon_2},\vec{\epsilon_1})\sim (\vec{\epsilon_3},\vec{\epsilon_3})\,, \\[0.4em]
(\vec{\epsilon_3},\vec{\epsilon_1})\sim (\vec{\epsilon_2},\vec{\epsilon_3})\,. \\[0.4em]
\end{array}
\end{equation}

Similarly, the transformation ${\cal P}^{(2)}$ defined by
\begin{equation}
    {\cal P}^{(2)} |x_1, x_2, x_3 \rangle = (-1)^{x_2} |x_3, x_2, x_1 \rangle \,,
\end{equation}
has the following action on the $k$-basis
\begin{equation}
    {\cal P}^{(2)} |k_1, k_2, k_3 \rangle =  (-1)^{N+k_1}|k_1, k_2, k_3 \rangle \,.
\end{equation}
Therefore
\begin{equation}{\cal P}^{(2)}\ :\ 
\left\{\begin{array}{l}
X_1\mapsto X_3\\[0.2em]
X_2\mapsto X_2\\[0.2em]
X_3\mapsto X_1
\end{array}\right.\ \ \ \ \text{and}\ \ \ \ \ \left\{\begin{array}{l}
H_1\mapsto H_1\\[0.2em]
H_2\mapsto H_2\\[0.2em]
H_3\mapsto H_3
\end{array}\right. \,,
\end{equation}
and similarly for the corresponding projectors.
It follows that the last two lines of \eqref{equivalence} also have the same entanglement entropy.

\subsubsection{Decoupled Tratnik case}\label{sec:decoupled}

\begin{figure}
	\centering
	\includegraphics[width=0.43\hsize]{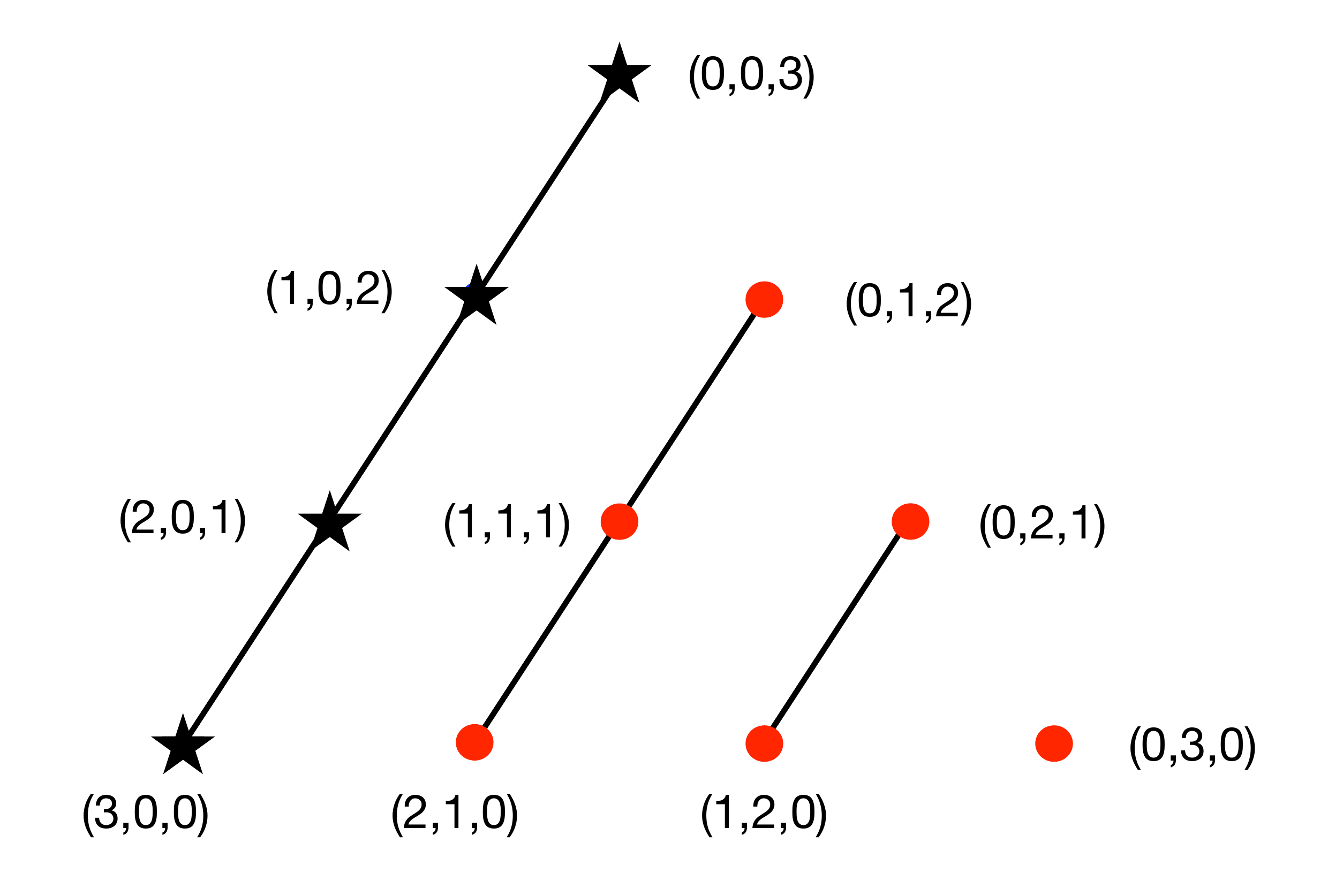}
	\caption{System with $N=D=3$, $\vec\alpha=\vec\epsilon_1$, $\vec\beta=\vec\epsilon_2$, and $L=0$. The vertices in the set $\setA$ \eqref{Sset} (the subsystem), which are labeled by their coordinates $(x_1,x_2,x_3)$, are denoted by stars, and lines connect vertices between which hopping is allowed by the Hamiltonian \eqref{Hamiltonian}. In this case, hopping is not allowed between the subsystem and its complement.}
	\label{fig:special}
\end{figure}

We observe that for the $D=3$ Tratnik case \eqref{Rtratnik} with $R_{12}=0$, the entanglement entropy exactly vanishes for a specific set of vectors $(\vec\alpha\,, \vec\beta)$, namely
\begin{equation}
    S_{vN}(\vec\alpha, \vec\beta, K, L, N) = 0\,, \qquad 
    \vec\alpha = \vec\epsilon_1\,, \quad \vec\beta=\vec\epsilon_2\,,
    \label{special}
\end{equation}
for all allowed values of $K, L, N$. This is consistent with the fact that, for this case, the subsystem does not interact with its complement, as illustrated in Fig. \ref{fig:special}.

More generally, it follows from the definitions \eqref{XH} that
\begin{equation}
    \left[ X_l \,, H_i \right] = \sum_{k=1}^DR_{il}R_{ik}a^{\dagger}_la_k-\sum_{j=1}^DR_{ij}R_{il}a^{\dagger}_ja_l \,.
    \label{commutespecial}
\end{equation}
This gives $0$ if $R_{il}=0$. We consider the following particular situation. Given a matrix $R$, we choose $\vec{\alpha}$ and $\vec{\beta}$ such that the only pairs $(\alpha_i,\beta_l)$ equal to $(1,1)$ correspond to coefficients $R_{il}=0$, that is,
\begin{equation}
\text{if}\ (\alpha_i,\beta_l)=(1,1)\ \ \ \ \ \text{then}\ \ R_{il}=0\ .
\label{conditioncommute}
\end{equation}
Then we deduce that in this case:
\begin{equation}
    \left[ \vec\alpha \cdot \vec H \,, \vec\beta \cdot \vec X  \right] = 0 \,.
    \label{commutespecial2}
\end{equation}
Now recall that $\pi_1$ is the projector on the sum of certain eigenspaces of $\vec\beta \cdot \vec X$, and as such it is a polynomial in $\vec\beta \cdot \vec X$. Similarly, $\pi_2$ is a polynomial in $\vec\alpha \cdot \vec H$. From (\ref{commutespecial2}), we have that $\pi_1$ and $\pi_2$ commute. Therefore they can be simultaneously diagonalized (with eigenvalues either $0$ or $1$) and thus $C = \pi_1\,  \pi_2\, \pi_1 = \pi_1\,  \pi_2$ has eigenvalues which are either $0$ or $1$. Therefore, the entanglement entropy vanishes in the particular situation (\ref{conditioncommute}).

\subsubsection{One-parameter case}\label{sec:1pD3}

\begin{figure}
	\centering
	\includegraphics[width=0.43\hsize]{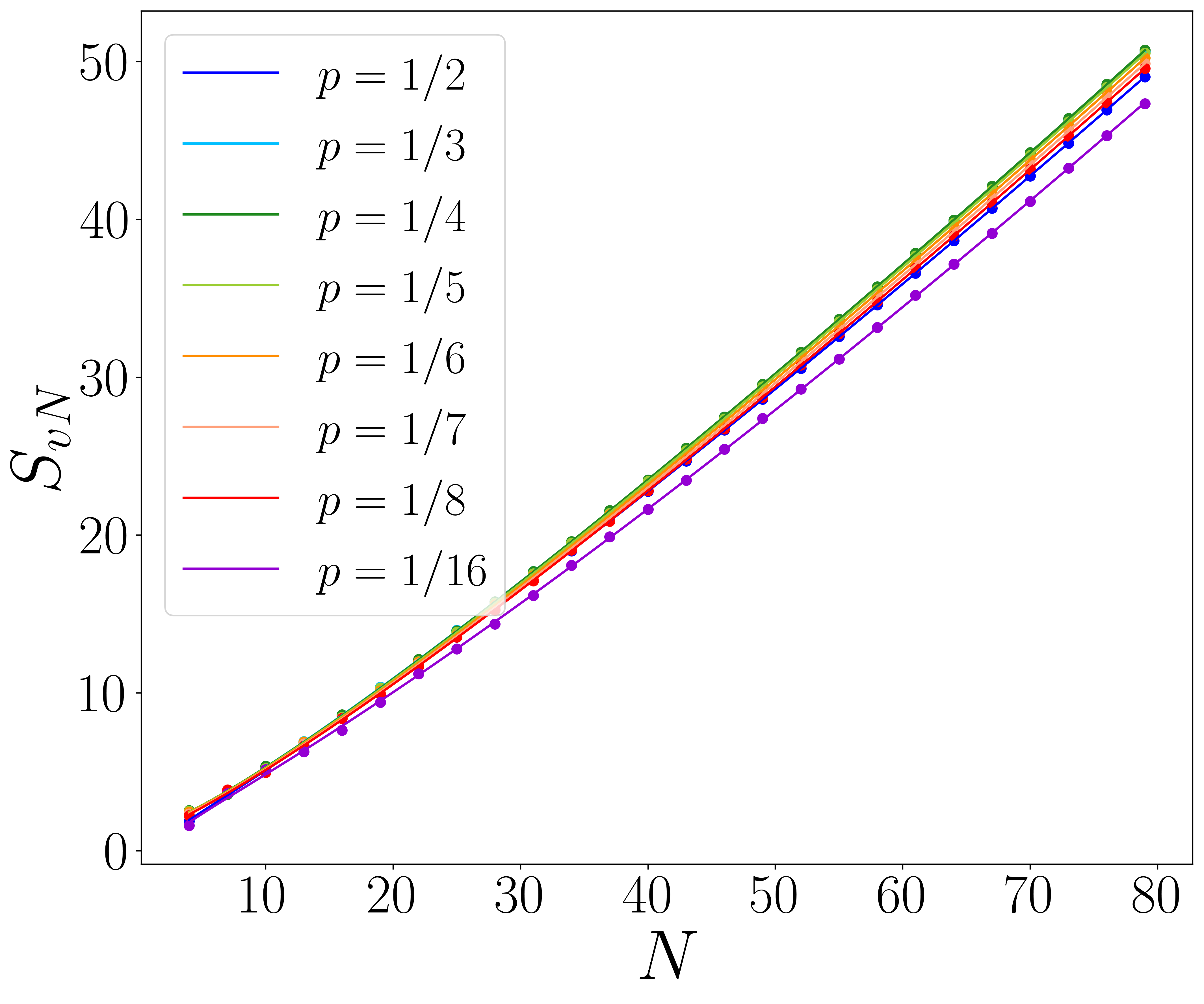}
		\includegraphics[width=0.43\hsize]{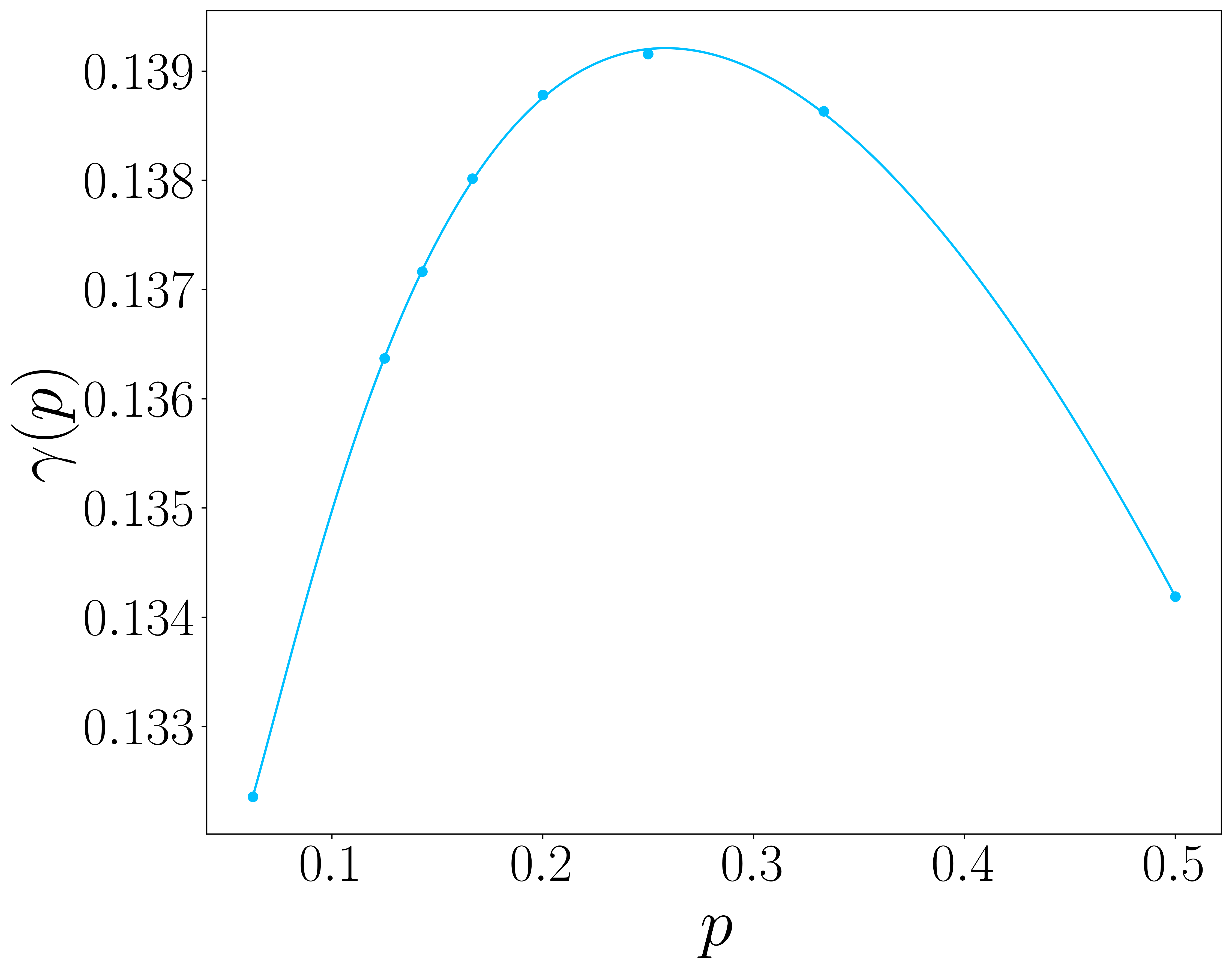}
	\caption{\textit{Left}: Entanglement entropy as a function of $N=3n+1$ for the one-parameter case with $D=3$, $L=K=N/3-1$ and $\vec{\alpha}=\vec{\beta}=\vec \epsilon_2$ for various values of $p$. The dots are obtained from direct diagonalization of the chopped correlation matrix, and the solid lines are fits of the form of Eq. \eqref{eq:SvnD3Trat}. \textit{Right}: Leading coefficient $\gamma$ in the expansion of Eq. \eqref{eq:SvnD3Trat} as a function of $p$. The dots are obtained from our numerical data, and the solid line serves as guides to the eye.}
	\label{fig:SvnOneparam}
\end{figure}

Let us now consider the one-parameter case for $D=3$, with Hamiltonian \eqref{Roneparam} whose eigenfunctions are given by \eqref{eigoneparam}, and let us set $\vec\alpha=\vec\epsilon_2$. (The case $\vec\alpha=\vec\epsilon_3$ is similar, while here $\vec\alpha=\vec\epsilon_1$  is trivial.) For $\vec\beta=\vec\epsilon_1$, the system becomes decoupled and has 0 entanglement entropy, as in Sec. \ref{sec:decoupled}. However, for $\vec\beta=\vec\alpha=\vec\epsilon_2$, the entanglement entropy is nonzero. As for the Tratnik case discussed in Sec. \ref{sec:D3Trat}, we observe plateaus if we consider all values of $N$. We impose $N=3n+1$, and again observe that the entropy exhibits a logarithmic violation of the area law, see the left panel of Fig. \ref{fig:SvnOneparam}. The leading coefficient $\gamma$ from the expansion \eqref{eq:SvnD3Trat} is displayed on the right panel of Fig. \ref{fig:SvnOneparam}. Similarly to the Tratnik case, we restrict ourselves to a numerical evaluation of this coefficient and the solid lines are simple guides to the eye.

A notable difference with the Tratnik case discussed of Sec. \ref{sec:D3Trat} is that $\gamma$ appears to depend only weakly on $p$. Most of the curves in the left panel of Fig. \ref{fig:SvnOneparam} are close to each other, as compared to the ones in the top panels of Fig. \ref{fig:SvnTratnik}. In particular, the curves for $p=1/3,1/4$ and $1/5$ are almost indistinguishable. A potential explanation for this behavior is the following. In the case of homogeneous free fermions on the square lattice, the scaling $S_{vN}\sim N \log N$ stems from the fact that the Hamiltonian in two dimensions can be expressed as a sum of Hamiltonians for one-dimensional chains via dimensional reduction \cite{chung2000density,murciano2020symmetry}. The entanglement entropy of a two-dimensional strip is exactly given as a sum of~$N$ entropies of one-dimensional chains at different filling fraction, and each of them scales as $\log N$. The dimensional reduction thus allows one to not only understand the logarithmic violation of the area law, but also to compute the coefficient of the leading term, as well as the sub-leading ones. In our model, considering the one-parameter case with $\vec \alpha= \vec \epsilon_2$ implies that the Hamiltonian \eqref{Hamiltonian} only contains hopping interactions along lines parallel to $\vec \epsilon_2-\vec \epsilon_3$, and hence can be seen as a sum of one-dimensional chains. The fact that the leading term of the entanglement entropy in one dimension is independent of~$p$ may help explain the weak dependence on $p$ here. However, because of the geometry of the problem, the computation of the leading coefficient $\gamma$ from one-dimensional chains appears more involved than on the square lattice, and we have not investigated that issue further.

\subsection[The case $D=4$]{The case $\boldsymbol{D=4}$}

In this section, we investigate the entanglement entropy for $D=4$. Based on our results for $D=2,3$, we expect the entropy to exhibits a logarithmic violation of the area law, and to scale as 
\begin{equation}
\label{eq:SvnD4}
    S_{vN}= \gamma \ N^2 \log N + \dots \,.
\end{equation}

We consider two particular cases, (i) the Tratnik case, see Eq. \eqref{Rtratnik4}, for which the eigenfunctions are given by \eqref{QTratnikD4}, and (ii) the one-parameter case discussed in Sec. \ref{sec:oneP}. Similarly as for the case $D=3$, we choose $L=K=N/4-1$. Without restrictions on the allowed values for $N$, we also observe plateau-like effects, that are only related to the geometry of the system and our choices of $L$ and $K$. Hence, we restrict ourselves to $N=4n+1$.

\begin{figure}
	\centering
	\includegraphics[width=0.43\hsize]{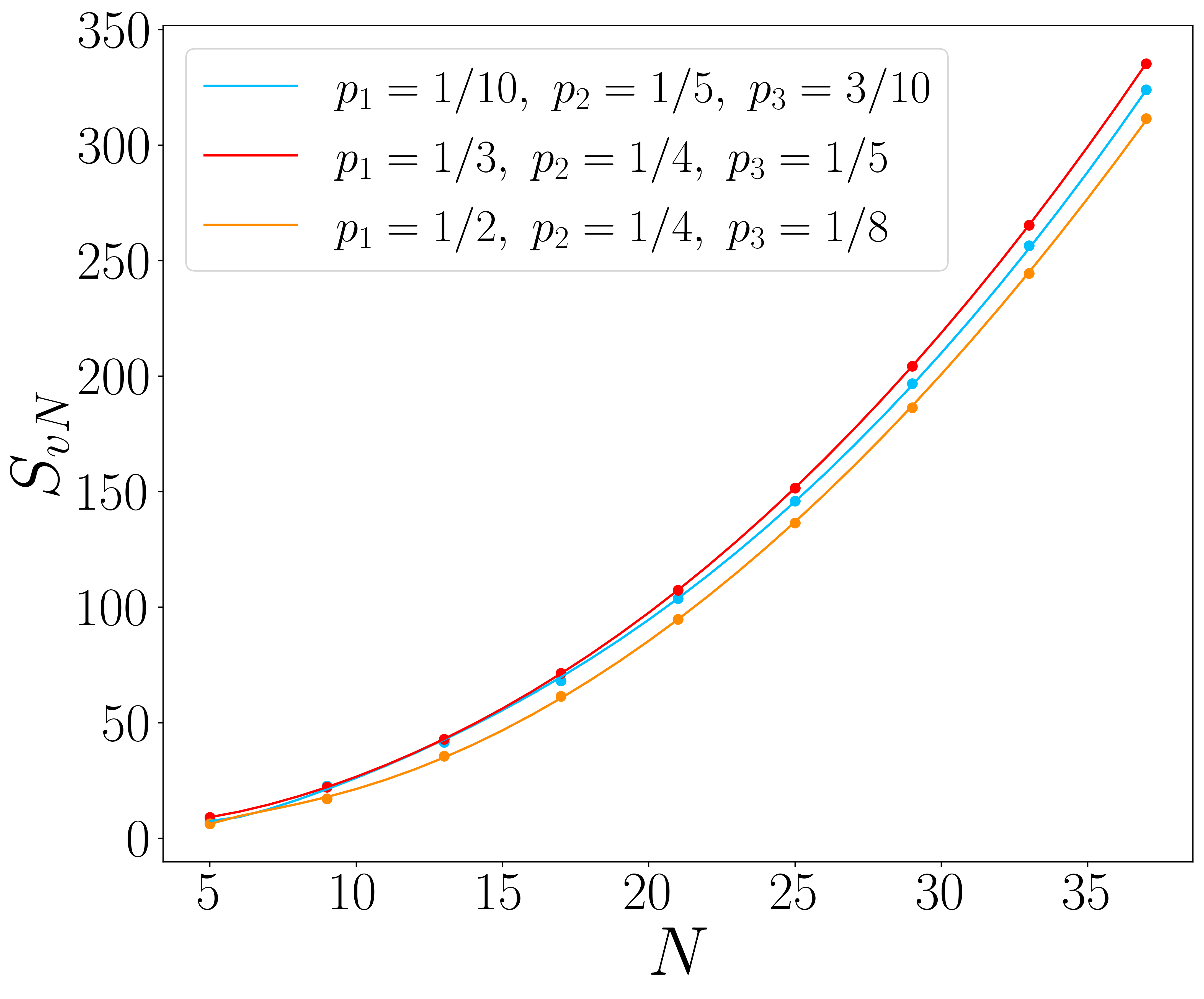}
	\includegraphics[width=0.43\hsize]{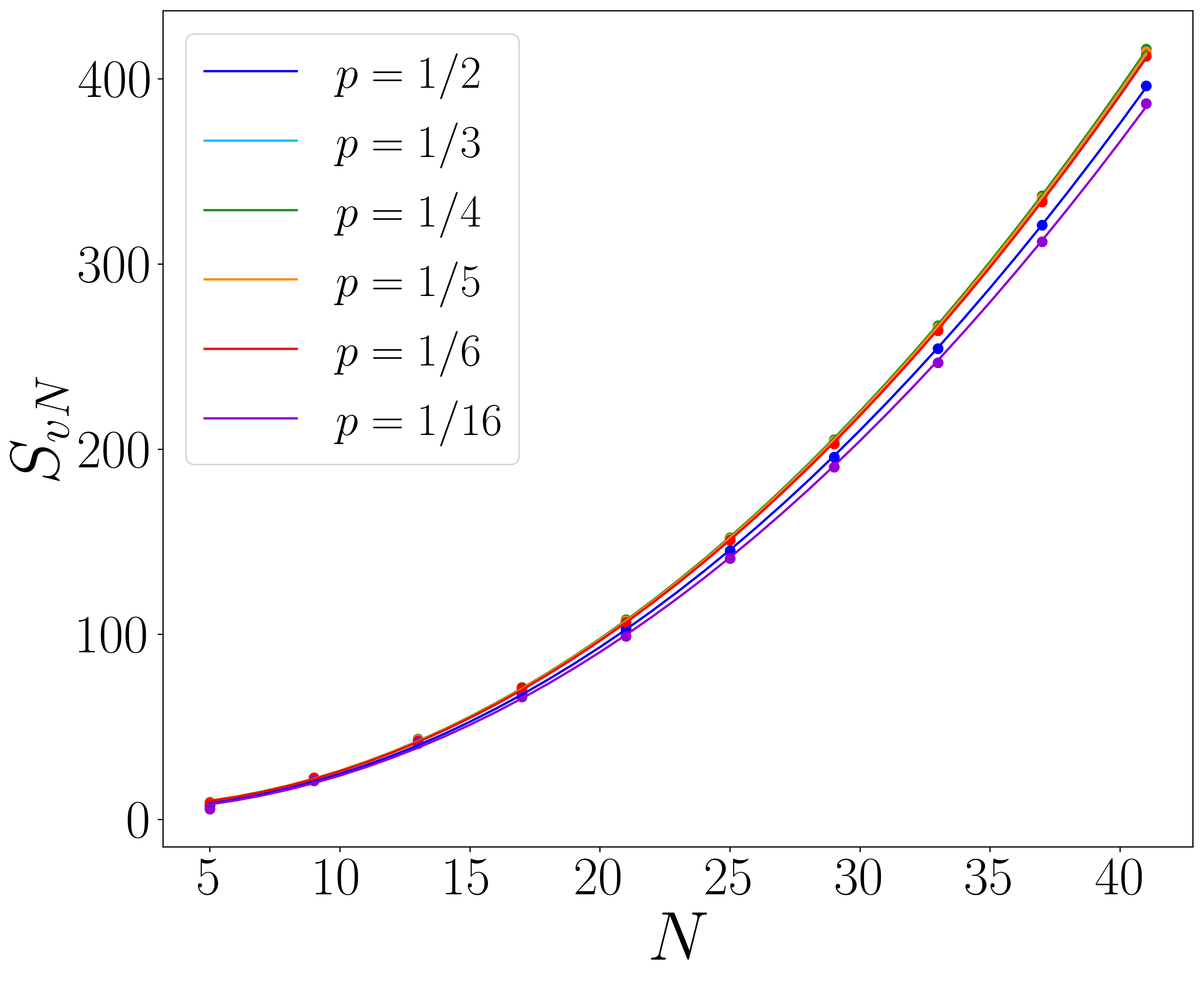}
	\caption{\textit{Left}: Entanglement entropy as a function of $N=4n+1$ for the Tratnik case with $D=4$, $L=K=N/4-1$ and $\vec{\alpha}=\vec{\beta} = \vec \epsilon_1$ for various values of $p_1,p_2,p_3$. \textit{Right}: Entanglement entropy as a function of $N=4n+1$ for the one-parameter case with $D=4$, $L=K=N/4-1$ and $\vec{\alpha}=\vec{\beta} = \vec \epsilon_3$ for various values of $p$. In both panels, the dots are obtained by direct diagonalization of the chopped correlation matrix, and the solid lines correspond to fits of the form of Eq. \eqref{eq:SvnD4}.}
	\label{fig:SvnD4}
\end{figure}

We display our numerical results in the left and right panels of Fig. \ref{fig:SvnD4} for cases (i) and (ii), respectively. The dots are obtained by direct diagonalization of the chopped correlation matrix, whereas the solid lines correspond to fits of the form of Eq. \eqref{eq:SvnD4}. 

For the one-parameter case, the curves for $p=1/3,1/4,1/5$ and $1/6$ are almost indistinguishable, and correspond to a leading term of $\gamma \simeq 0.06$. Our heuristic explanation for this phenomenon is similar as the one we discuss in Sec. \ref{sec:1pD3} for the one-parameter case for $D=3$. 

\section{Conclusion and outlook}\label{sec:end}

In this paper, we introduced and solved a new inhomogeneous model of free fermions in arbitrary spatial dimensions. The model coincides with the Krawtchouk chain \cite{Crampe:2019upj} in the one-dimensional case, and its solution in arbitrary dimensions relies on multivariate Krawtchouk polynomials. Moreover, using bispectral properties of the problem, we constructed an operator that commutes with the chopped correlation matrix and provides a natural extension of the Heun operator obtained in \cite{Crampe:2019upj} to higher dimensions. 

We investigated the entanglement properties of half-filled eigenstates of the Hamiltonian \eqref{Hamiltonian} for $D=2,3,4$, which correspond to systems in one, two and three spatial dimensions, respectively. We found that the entanglement entropy scales as 
\begin{equation}
   S_{vN} = \gamma \ N^{D-2}\log N + \dots \, , 
\end{equation}
and thus exhibits a logarithmic violation of the area law. For $D=2$, the leading coefficient is $\gamma=1/6$, in agreement with previous results obtained in the framework of curved-space CFT \cite{FA21}. We also observed and conjectured the exact form of sub-leading oscillations, see Eq. \eqref{eq:mp}, but an analytical proof is beyond the scope of this paper. Crucially, the frequencies of those oscillations depend on the model through the parameter $p$. This property is in stark contrast with known sub-leading oscillations in homogeneous systems, which depend only on the filling fraction, and represents one of the main results of this paper. For $D=3,4$, the leading coefficient $\gamma$ depends on the parameters of the model, and this dependence is much weaker for the one-parameter cases. We restricted ourselves to numerical evaluations of this coefficient.

There are many open problems that are worth investigating in the future. A first one would be to provide an analytical proof for the conjecture of Eqs. \eqref{eq:Svn1dConj}, \eqref{eq:mp}, and hence extend the results of Ref. \cite{fagotti2011universal} to inhomogeneous systems. However, we expect this problem to be very challenging, since the inhomogeneity of the model breaks the Toeplitz and Hankel structure of the correlation matrix. Another complementary idea is to use curved-space CFT methods to understand the sub-leading oscillations, similarly to the Rainbow chain \cite{rodriguez2017more}, as well as the leading terms for $p\neq 1/2$. In parallel to analytical results, we believe that the model-dependent sub-leading oscillations deserve further investigations in the context of other inhomogeneous chains. For instance, preliminary results suggest that a similar phenomenon occurs for the anti-Krawtchouk chain \cite{GVYZ2016}. It would also be interesting to better understand the dependence of $\gamma$ on the parameters of the model for $D=3,4$, either via analytical computations or with field-theoretical arguments in higher dimensions. Another intriguing question is whether one could use the Heun operator to compute entanglement entropies faster and with higher precision than via direct diagonalization of the chopped correlation matrix. Preliminary results seem to suggest so. Moreover, the Heun operator could also be used to derive analytical results for the entanglement entropy via Bethe ansatz techniques \cite{bernard2021heun}. Other natural extensions of our work include the investigation of inhomogeneous models based on other representations of $su(D)$ and on graphs of P- and Q- polynomial association schemes \cite{Bannai1984AlgebraicCI, brouwer2012distance}. The entanglement entropy of free fermions on the celebrated Hamming scheme was found to be related to the entanglement in Krawtchouk chains \cite{bernard2021entanglement}, and similarly the entanglement of free fermions on ordered Hamming graphs \cite{miki2019quantum} is expected to be related to the entanglement in the higher-dimensional models introduced in this paper.

\paragraph{Acknowledgements}

PAB holds an Alexander-Graham-Bell scholarship from the Natural Sciences and Engineering Research Council of Canada (NSERC). NC and LPA are supported by the international research project AAPT of the CNRS and the ANR Project AHA ANR-18-CE40-0001.
RN gratefully acknowledges financial support from a CRM--Simons professorship, and the warm hospitality extended to him at the Centre de Recherches Math\'ematiques (CRM) during his visit to Montreal. GP holds a CRM--ISM postdoctoral fellowship and acknowledges support from the Mathematical Physics Laboratory of the CRM. The research of LV is supported by a Discovery Grant from NSERC. GP thanks Cl\'ement Berthiere, Riccarda Bonsignori and Juliette Geoffrion for useful discussions. 

\bibliographystyle{utphys}

\providecommand{\href}[2]{#2}\begingroup\raggedright\endgroup

\end{document}